\documentclass[aps,prb,twocolumn,nofootinbib,superscriptaddress,10pt,floatfix]{revtex4-2}
\usepackage{amsmath,amsthm,amsfonts,amssymb,amscd,bbold,bm}
\usepackage[T1]{fontenc}
\usepackage{physics}
\usepackage{xcolor}
\usepackage{graphicx}
\usepackage{tabularx}
\usepackage{subfigure}
\usepackage{float}
\usepackage{ulem}
\usepackage{hyperref}
\usepackage{changepage}

\usepackage{comment}

\begin{document}

\title{Chiral classical and quantum acoustics with hole-spin qubits}
\author{Zhanning Wang}
\author{Yongtao Li}
\affiliation{Quantum Advanced Research Center (QuARC) - Consejo Superior de Investigaciones Cient{\'i}ficas (CSIC)}
\affiliation{Instituto de Ciencia de Materiales de Madrid (ICMM) - Consejo Superior de Investigaciones Cient{\'i}ficas (CSIC) - Sor Juana In{\'e}s de la Cruz 3, 28049 Madrid, Spain}

\author{Nelson E. Rivas}
\affiliation{Instituto de Ciencia de Materiales de Madrid (ICMM) - Consejo Superior de Investigaciones Cient{\'i}ficas (CSIC) - Sor Juana In{\'e}s de la Cruz 3, 28049 Madrid, Spain}

\author{Gonzalo Garc\'ia}
\author{Irene Castro}
\author{Rub\'en Seoane Souto}
\author{Daniel Ramos}
\author{Jos\'e C. Abadillo-Uriel}
\email{jc.abadillo.uriel@csic.es}
\affiliation{Quantum Advanced Research Center (QuARC) - Consejo Superior de Investigaciones Cient{\'i}ficas (CSIC)}
\affiliation{Instituto de Ciencia de Materiales de Madrid (ICMM) - Consejo Superior de Investigaciones Cient{\'i}ficas (CSIC) - Sor Juana In{\'e}s de la Cruz 3, 28049 Madrid, Spain}
\date{\today}

\begin{abstract}
Gate-defined hole spins combine strong spin-orbit coupling with exceptional strain sensitivity, making surface acoustic waves a natural route to remote, phase-coherent control.
We show that counterpropagating surface acoustic waves can differ in coupling strength and in whether they predominantly drive spin rotations or modulate the qubit frequency.
We call the transverse transition-strength imbalance spin-acoustic chirality and show that it is tunable through gate-controlled reshaping of the dot confinement.
Combining a multiband Luttinger-Kohn Hamiltonian with the Bir-Pikus description of strain and 3D piezoelectric finite-element simulations, we obtain the acoustic $g$-matrix modulation governing coherent SAW driving.
Quantizing the same strain-mediated interaction yields the corresponding single-phonon coupling, and a generalized anisotropic Rabi interaction with co-rotating, counter-rotating, and longitudinal components, linking classical chirality to quantum anisotropy and directional one-phonon emission.
Magnetic field orientation and dot shape tune the acoustic chirality and operator selectivity.
We illustrate this framework through bichromatic coherence protection and phonon-mediated heralded Bell-state initialization.
\end{abstract}
\maketitle

\section{Introduction}
\label{Sec_1_introduction}
Gate-defined semiconductor spin qubits offer a direct route to compact quantum processors. The qubit, its tunnel couplings, and its readout geometry are all set by lithographic electrodes in scalable semiconductor heterostructures~\cite{Loss1998,Kloeffel2013,Burkard2023}. Planar Ge hole spins are especially promising for nanometer-scale confinement supporting dense two-dimensional (2D) layouts, while valence band spin-orbit coupling (SOC) enables fast all-electrical control without micromagnets~\cite{Scappucci2020,Hendrickx2021}. Coherent single- and two-qubit operations have been demonstrated at optimized working points, and recent experiments have extended control to multidot devices and small arrays~\cite{Froning2021,Wang2022,Hendrickx2024,Bassi2025}. As these arrays grow, distributing phase-coherent high-frequency control without routing a dedicated line to every dot becomes a systems-level challenge.

Surface acoustic waves (SAWs) are promising to address such a challenge.
A SAW is launched remotely by an interdigital transducer (IDT) and propagates at a GHz frequency that matches a spin transition~\cite{Chu2026}.
Its micron-scale wavelength is comparable to the spacing between the quantum dots.
It therefore carries a deterministic phase reference across spatially separated sites~\cite{White1965,Delsing2019}.
Because the mode is confined near the surface, it can generate appreciable dynamic strain in a shallow quantum well.
A single launched tone can act as a shared high-frequency field for multiple qubits, reducing the need to route a separate high-frequency drive line to every site, in analogy with global-control concepts for donor qubits~\cite{Kane1998}.
Furthermore, SAWs can be used to generate entanglement between distant qubits~\cite{Schuetz2015}.

Ge hole spins are well suited to such SAW-based control because their Zeeman response and spin-orbit structure are strain sensitive, producing sizable qubit driving and modulation~\cite{Abadillo2023}.
Although our calculations use Ge, the same strain-mediated mechanism applies qualitatively to Si hole spins, with material- and confinement-dependent coupling strengths.
Static gates can additionally maintain each qubit near a favorable operating point and locally tune its response to the shared SAW through dc control~\cite{wang2024operating,rimbach2025gapless,martinez2026disorder}, thereby combining global acoustic delivery with local electrostatic programmability.
The same acoustic mode can furthermore be treated either as a coherently driven classical field or as a quantized excitation, directly linking distributed qubit control to cavity and traveling-phonon quantum acoustics.
\begin{figure*}[htbp!]
\centering
\includegraphics[width=\linewidth]{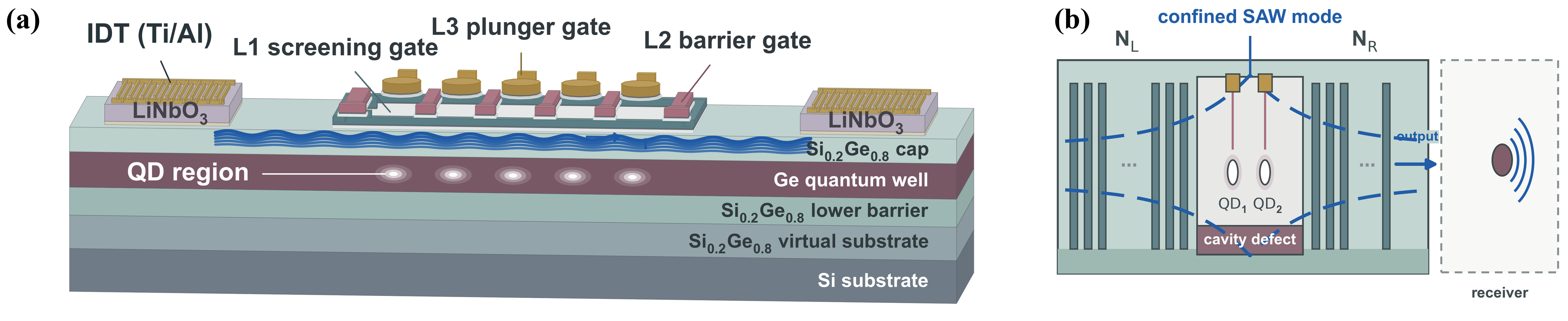}
\caption{(a) Schematic of a SAW traveling through a multidot device, comprising two LiNbO$_3$-based IDTs  with voltages $V_\text{IDT}$ integrated with a layered Si/SiGe/Ge heterostructure.
In our simulations, quantum dots are defined within the Ge quantum well and controlled by a screening gate, six barrier gates and five plunger gates.
The blue wavefronts indicate the SAW launched by the IDTs. (b) Schematic of a shared phononic-crystal (PnC) cavity containing two quantum dots within the central defect region. The two dots couple to the same confined SAW mode. The left and right mirrors contain $N_{\text{L}}$ and $N_{\text{R}}$ cells, respectively, with $N_{\text{L}}>N_{\text{R}}$. They confine the mode and preferentially direct acoustic output through the shorter right mirror.}
\label{Fig_1}
\end{figure*}

Much of this quantum-acoustic toolbox has been developed with superconducting circuits, which now generate, transfer, and entangle phonon states~\cite{Gustafsson2014,Satzinger2018,Bienfait2019,Dumur2021}.
It has also been developed with phononic crystals (PnCs), which shape the acoustic local density of states through engineered band gaps~\cite{MacCabe2020,Odeh2025}.
Besides, thin-film lithium niobate enables piezo-optomechanical transduction and integration with superconducting circuits~\cite{Mirhosseini2020,Jiang2020,Sahu2025}, and SAWs coherently drive optically addressable color centers~\cite{Golter2016,Maity2020,HernandezMinguez2020}.
Propagating phonons have distinctive kinematics: their velocity lies five orders of magnitude below that of light.
A GHz mode therefore has a micron-scale wavelength and a device-scale travel time.
The elliptical polarization of a Rayleigh wave also reverses handedness with the propagation direction.
These features give access to chiral interactions, non-Markovian propagation delays, and giant-atom physics~\cite{Bliokh2015,Yuan2021,Soro2022}, phenomena that gate-defined spin qubits have yet to access.

Research on acoustic control of confined holes is beginning to bridge these developments.
Microscopic studies have established electric dipole spin resonance, g-tensor modulation, and strain effects in valence-band qubits~\cite{Bulaev2007,Wang2021,Bosco2021b,Wang2022,Martinez2022,Abadillo2023,Rodriguez2023,Sarkar2025}.
Recent work has modeled SAW-induced strain in Ge-based heterostructures and shown that the SAW-driven Rabi frequency depends strongly on magnetic field orientation and confinement~\cite{Yonjali2025,YuanHsiao2025}.
In parallel, proposals have considered PnC and long-range links for hole spins, including cavity designs in high-purity germanium and deformation-potential coupling through engineered waveguides~\cite{Mei2025,Myronov2026}.
A unified, microscopically informed description of how counterpropagating acoustic modes project onto a gate-defined hole spin qubit is still lacking.

In this work, we describe a gate-defined Ge hole spin coupled to Rayleigh-type acoustic modes using the microscopic six-band Luttinger-Kohn Hamiltonian together with the Bir-Pikus terms for strain (LKBP model in the following)~\cite{Luttinger1955,Luttinger1956}.
We derive a direction-resolved acoustic g-matrix modulation that provides a unified description of coherent SAW driving, longitudinal modulation, and coupling to quantized cavity or traveling phonons.
We show that electrostatic gating and magnetic field orientation control the overall coupling strength, the directional chirality of the transverse coupling, and the balance between longitudinal modulation and co-rotating transverse driving.
We then quantize the strain field and resolve the resulting spin-phonon interaction into co-rotating, counter-rotating, and longitudinal components.
As applications of this framework, we identify acoustically dressed operating points with enhanced coherence against electrical fluctuations and investigate heralded Bell-state initialization in a PnC mode.
Together, these results establish electrically programmable spin-acoustic couplings for hole-spin control and interconnection.

\section{Model and theory}
\label{Sec_2_model}

We describe the coupling between a Rayleigh-type SAW and a Ge hole-spin qubit as an acoustic modulation of the qubit $g$-matrix~\cite{Venitucci2018GMatrix} in the device geometry of Fig.~\ref{Fig_1}(a).
The qubit is the lowest Zeeman-split doublet of a confined valence-band hole.
The acoustic mode, launched by lithium-niobate IDTs, propagates in the quantum-well plane along $\pm\hat{\bm{e}}$.
The acoustic axis is $\hat{\bm{e}}=(\cos\theta_{\text{SAW}},\sin\theta_{\text{SAW}},0)$, where $\theta_{\text{SAW}}$ is the angle between the lab-frame $x$ direction and $+\hat{\bm{e}}$.
The wave vector is $\bm{q}_\xi=\xi k\hat{\bm{e}}$, with $\xi=+1$ and $-1$ denoting propagation along $+\hat{\bm{e}}$ and $-\hat{\bm{e}}$, respectively.

Projecting the full six-band LKBP Hamiltonian onto the lowest Zeeman-split doublet (Appendix~\ref{App:QubitProjection}) yields:
\begin{equation}
\label{Eq:qubit_hamiltonian}
H_{\text{Q},\xi}(t) = \frac{\mu_{\text{B}}}{2}\bm{\sigma} \cdot [\tilde{g}_0 + \delta\tilde{g}_\xi(t)]\bm{B} = \frac{\hbar}{2}[\bm{\Omega}_0 + \delta\bm{\Omega}_\xi(t)] \cdot \bm{\sigma} \,.
\end{equation}
Here $\tilde{g}_0$ is the static $g$-matrix and $\delta\tilde{g}_\xi(t)$ its SAW-induced modulation.
The $g$-matrix maps the laboratory magnetic field $\bm{B}$ onto the Larmor vector in the qubit Pauli space, with $\bm{\Omega}_0 = (\mu_{\text{B}}/\hbar)\tilde{g}_0\bm{B}$, $\omega_q = |\bm{\Omega}_0|$, and $\hat{\bm{n}} = \bm{\Omega}_0/\omega_q$. The longitudinal component along $\hat{\bm{n}}$ modulates the qubit splitting; at resonance, the co-rotating transverse component drives spin rotations. The static $g$-matrix already includes the device-specific static strain and electrostatic confinement, which fix the heavy-hole-light-hole (HH-LH) composition of the qubit states.

The acoustic strain enters the same projection through the linear Bir-Pikus coupling to the material deformation potentials $a_v$, $b_v$, and $d_v$.
To leading order in the HH-LH mixing~\cite{Piot2022,Abadillo2023}, this coupling gives the following corrections to the $g$-matrix:
\begin{subequations}
\label{Eq:g_corrections}
\begin{align}
\delta\tilde{g}_{xx} &= \delta\tilde{g}_{yy} = \frac{6b_v\kappa}{E_{\text{LH}}-E_{\text{HH}}}\expval**{\epsilon_{yy}-\epsilon_{xx}} \,, \\
\delta\tilde{g}_{xy} &= -\delta\tilde{g}_{yx} = \frac{4\sqrt{3}\kappa d_v}{E_{\text{LH}}-E_{\text{HH}}}\expval**{\epsilon_{xy}} \,, \\
\delta\tilde{g}_{zx} &= -\frac{4\sqrt{3}\kappa d_v}{E_{\text{LH}}-E_{\text{HH}}}\expval**{\epsilon_{xz}} \,, \\
\delta\tilde{g}_{zy} &= -\frac{4\sqrt{3}\kappa d_v}{E_{\text{LH}}-E_{\text{HH}}}\expval**{\epsilon_{yz}} \,.
\end{align}
\end{subequations}
Here $\expval{\cdots}$ denotes the orbital average over the dot wave function, and $\kappa$ is the valence band Zeeman parameter.
The HH-LH splitting $E_{\text{LH}}-E_{\text{HH}}$ is set by the static strain and the vertical confinement. Here $\epsilon_{\mu\nu}$ denotes the $(\mu,\nu)$ component of the strain tensor at the quantum dot.

The in-plane normal-strain anisotropy $\epsilon_{yy}-\epsilon_{xx}$ modulates the diagonal in-plane entries of the $g$-matrix.
The in-plane shear strain $\epsilon_{xy}$ generates an antisymmetric contribution to the in-plane $g$-matrix.
The sagittal shear components $\epsilon_{xz}$ and $\epsilon_{yz}$ generate the $zx$ and $zy$ entries, respectively.
The latter convert an in-plane magnetic field into an out-of-plane component of the acoustic Larmor vector. 
Leading-order relations in Eqs.~\eqref{Eq:g_corrections} provide physical intuition; all numerical results use the complete LKBP model.

An ideal Rayleigh wave is polarized in the sagittal plane, namely the plane spanned by the propagation direction $\hat{\bm{e}}$ and the growth direction $\hat{\mathbf{z}}$.
A material point near the surface  follows an ellipse with in-plane and vertical motion.
At the qubit, this motion predominantly produces a normal strain along the propagation axis, $\epsilon_{\parallel\parallel}$, a vertical normal strain, $\epsilon_{zz}$, and a sagittal shear strain, $\epsilon_{\parallel z}$.
Their relative phases are fixed by the elastic eigenmode and retained in the finite element calculation.

We use the basis $\{\hat{\bm{e}},\hat{\bm{e}}_\perp,\hat{\mathbf{z}}\}$, with $\hat{\bm{e}}_\perp$ perpendicular to both $\hat{\bm{e}}$ and $z$.
In this basis, the strain at the dot has the following approximate schematic form:
\begin{equation}
\label{Eq:rayleigh_strain_schematic}
\epsilon_\xi \sim \mqty[\epsilon_{\parallel\parallel} & 0 & \xi\epsilon_{\parallel z} \\ 0 & 0 & 0 \\ \xi\epsilon_{\parallel z} & 0 & \epsilon_{zz}]_{(\parallel,\perp,z)} \,.
\end{equation}
The entries denote harmonic amplitudes in the frame set by $\hat{\bm{e}}$, its in-plane normal, and $\hat{\mathbf{z}}$.
Reversing the propagation direction through $\xi\to -\xi$ leaves the normal-strain sector unchanged but reverses the sagittal shear, which is the strain-level consequence of reversing the handedness of the local elliptical motion.

Rotating into the crystal frame gives the combinations that enter Eq.~\eqref{Eq:g_corrections}:
\begin{subequations}\label{Eq:rayleigh_to_crystal}
\begin{align}
\epsilon_{yy}-\epsilon_{xx} &= -\cos(2\theta_{\text{SAW}})\epsilon_{\parallel\parallel} \,, \\
\epsilon_{xy} &= \frac{1}{2}\sin(2\theta_{\text{SAW}})\epsilon_{\parallel\parallel} \,, \\
\left(\epsilon_{xz},\epsilon_{yz}\right)_\xi &= \xi\epsilon_{\parallel z}\left(\cos\theta_{\text{SAW}},\sin\theta_{\text{SAW}}\right) \,.
\end{align}
\end{subequations}
Together, Eqs.~\eqref{Eq:g_corrections} and~\eqref{Eq:rayleigh_to_crystal} determine how the two strain sectors contribute to directionality.
The in-plane strain sector feeds the in-plane $g$-matrix and is unchanged under propagation reversal.
The sagittal shear feeds $\delta\tilde{g}_{zx}$ and $\delta\tilde{g}_{zy}$ and changes sign under propagation reversal.
After multiplication by the magnetic field and projection onto the co-rotating transverse component, the two strain sectors can interfere constructively for one propagation direction and destructively for the other.
The full LKBP calculation supplements Eq.~\eqref{Eq:g_corrections} with the $\epsilon_{zz}$ contributions to $P_\epsilon$ and $Q_\epsilon$. These contributions shift the band energies and HH-LH splitting. The full calculation retains all six strain components from the 3D piezoelectric FEM in Appendix~\ref{App:FiniteElementSimulations}.

For a monochromatic SAW of angular frequency $\omega$, we write the real-time modulation as:
\begin{equation}\label{Eq:gmatrix_quadratures}
\delta\tilde{g}_\xi(t) = \delta\tilde{g}^{(c)}_\xi\cos\omega t + \delta\tilde{g}^{(s)}_\xi\sin\omega t \,.
\end{equation}
Here $\delta\tilde{g}^{(c)}_\xi$ and $\delta\tilde{g}^{(s)}_\xi$ are real $g$-matrix quadratures obtained from the projected FEM field.
Together, these two quadratures encode the relative phases of all strain components.
The longitudinal and transverse responses then follow as:
\begin{small}
\begin{subequations}\label{Eq:gmatrix_fTL}
\begin{align}
f_{L,\xi} &= \frac{\mu_{\text{B}}}{2\pi\hbar}\left\{\left[\hat{\bm{n}}\cdot\delta\tilde{g}^{(c)}_\xi\bm{B}\right]^2+\left[\hat{\bm{n}}\cdot\delta\tilde{g}^{(s)}_\xi\bm{B}\right]^2\right\}^{1/2} \,, \\
f_{T,\xi} &= \frac{\mu_{\text{B}}}{4\pi\hbar}\left|\left(\delta\tilde{g}^{(s)}_\xi\bm{B}\right)\times\hat{\bm{n}}-\left[\left(\delta\tilde{g}^{(c)}_\xi\bm{B}\right)\times\hat{\bm{n}}\right]\times\hat{\bm{n}}\right| \,.
\end{align}
\end{subequations}   
\end{small}
Here $f_{L,\xi}$ is the peak amplitude of the SAW-induced qubit frequency modulation, and $f_{T,\xi}$ is, at resonance, the Rabi frequency in the rotating-wave approximation.
The second line combines the two $g$-matrix quadratures to obtain the co-rotating transverse component that determines the resonant Rabi frequency~\cite{golovach2006electric,Abadillo2023}.

For the two counter-propagating driven fields normalized to equal acoustic energy flux, the directional contrast of the transverse coupling is:
\begin{equation}\label{Eq:chirality}
\chi = \frac{f_{T,+}^2-f_{T,-}^2}{f_{T,+}^2+f_{T,-}^2} \,,
\end{equation}
which we refer to simply as the chirality.
The two directions are compared at equal acoustic energy flux, and the squared rates enter because transition strengths scale with the squared transverse coupling.

A nonzero $\chi$ comes from strain fields producing different co-rotating $g$-matrix modulations at a fixed magnetic field and working point.
Propagation reversal leaves the in-plane strain contribution unchanged but reverses the sagittal-shear contribution.
The two contributions therefore add in one direction and subtract in the other.
Their relative phase and their projection relative to the Larmor axis set the sign and magnitude of $\chi$.
In-plane magnetic field angle, dot anisotropy, and acoustic axis tune that projection through the qubit eigenstates, the Larmor axis, and active entries of $\delta\tilde{g}_\xi$.
Together, these controls tune the acoustic response to the incoming propagation direction.

\section{Programmable spin-acoustic chirality}
\label{Sec_3_results}
Section~\ref{Sec_2_model} showed that the normal and sagittal-shear components of a Rayleigh-type wave modify different entries of the qubit $g$-matrix and transform differently under propagation reversal.
We now determine whether this mechanism produces experimentally useful operator and direction selectivity in the realistic device field, and whether that selectivity can be controlled through the qubit working point.

For the results below, the complete complex strain tensor at the Ge quantum well is obtained from the 3D piezoelectric finite element model at $f_{\text{SAW}} = 2$~GHz and is projected onto the multiband hole states. The FEM reference field is calculated directly for $\theta_{\text{SAW}}=0^\circ$ and rotated to each nonzero propagation angle. The finite element implementation is given in Appendix~\ref{App:FiniteElementSimulations}. We model the quantum dot as a disorder-free device with ideal harmonic confinement.

We have numerically verified that vertical components of the magnetic fields do not yield a finite spin-SAW coupling, as expected from Eqs.~\eqref{Eq:g_corrections}. For an in-plane magnetic field, we use $\bm{B} = B_{\text{res}}\hat{\bm{b}}$, with $\hat{\bm{b}} = \left(\cos\theta_B,\sin\theta_B,0\right)$.
Furthermore, we can define the Larmor axis as $\hat{\bm{n}} = \tilde{g}_0 \hat{\bm{b}}/|\tilde{g}_0\hat{\bm{b}}|$, such that $B_{\text{res}} = hf_{\text{SAW}}/(\mu_{\text{B}}|\tilde{g}_0\hat{\bm{b}}|)$.
The static $g$-matrix determines both the field magnitude required for resonance and the Larmor axis relative to which the acoustic modulation is classified as transverse or longitudinal.
The dependence of $\tilde{g}_0$ on the confinement parameters is left implicit here and is discussed explicitly in Sec.~\ref{Sec_3_3_confinement}.
At every magnetic field and confinement working point we rediagonalize the static LKBP Hamiltonian, recalculate $\tilde{g}_0$, $\hat{\bm{n}}$, and $\delta\tilde{g}_\xi$, and adjust $B_{\text{res}}$ so that $\omega_q = \omega_{\text{SAW}}$ is maintained.

\subsection{Direction-dependent transverse and longitudinal coupling}
\label{Sec_3_1_transverse_longitudinal}
\begin{figure}[htbp!]
\centering
\includegraphics[width=\linewidth]{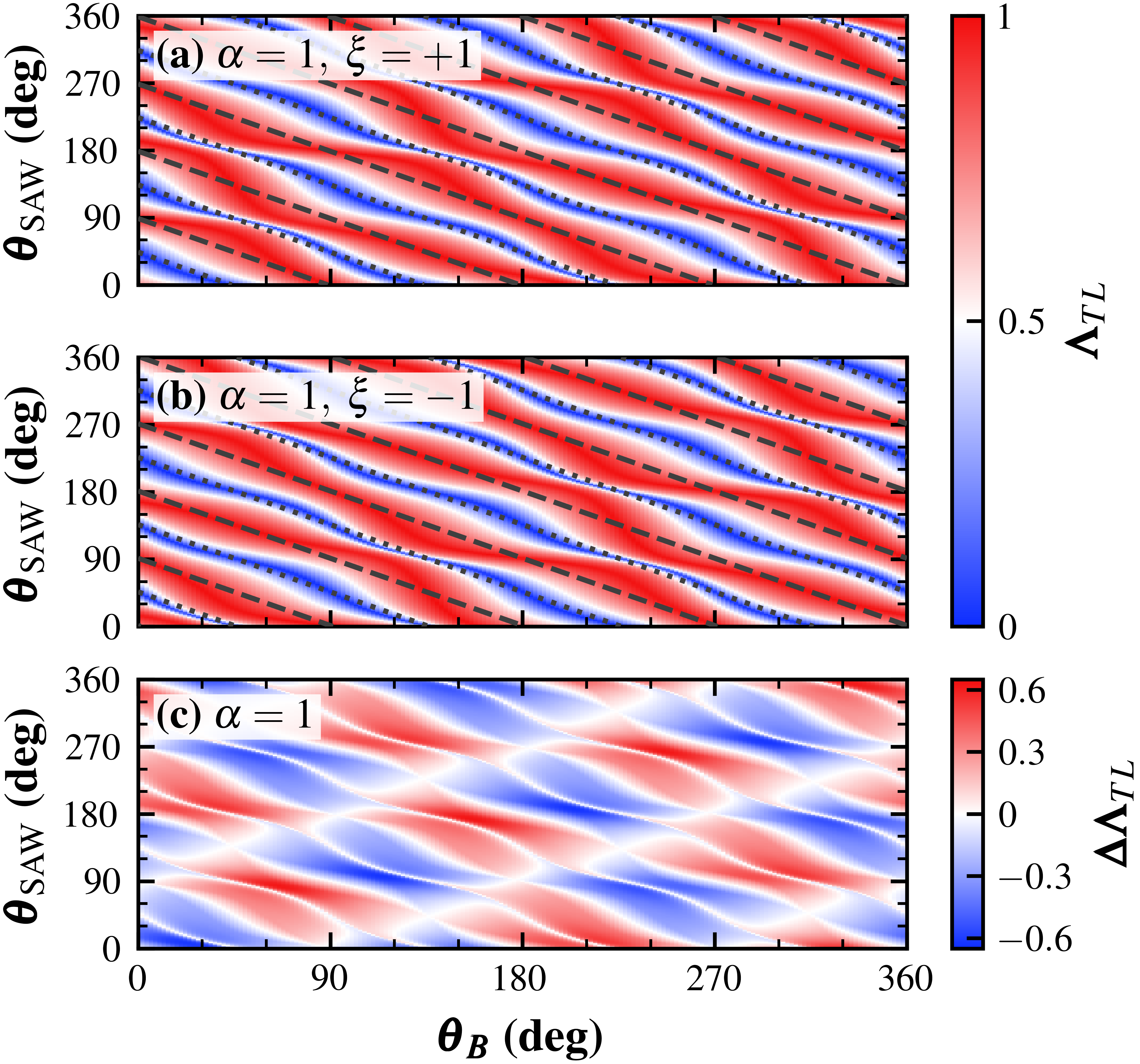}
\caption{Longitudinal/co-rotating-transverse composition of the acoustic $g$-matrix modulation for the two propagation branches.
The color map shows the transverse-longitudinal mixing parameter $\Lambda_{\text{TL},\xi}$ defined in Eq.~\eqref{Eq:lambda_TL}.
It is plotted as a function of the acoustic axis $\theta_{\text{SAW}}$ and the magnetic field angle $\theta_B$.
The calculation uses the $2$~GHz FEM strain field for a circular quantum dot, $\alpha=1$.
Panels (a) and (b) correspond to $\xi=+1$ and $\xi=-1$, respectively.
The straight dashed and dotted grey lines in (a) and (b) are guides to the eye for this dominant fitted fourfold harmonic.
The weaker harmonics produce the deviations of the numerical texture from these guides.
(c) shows the branch difference $\Delta\Lambda_{\mathrm{TL}}=\Lambda_{\mathrm{TL},+}-\Lambda_{\mathrm{TL},-}$. 
Blue, $\Lambda_{\text{TL},\xi}=0$, denotes a purely transverse response and red, $\Lambda_{\text{TL},\xi}=1$, a purely longitudinal one.}
\label{Fig_2}
\end{figure}

Depending on the projection of its acoustic Larmor vector, a Rayleigh-type SAW can predominantly drive spin rotations or modulate the qubit precession frequency. Therefore, we evaluate the possibility of having the $\xi=\pm 1$ branches addressing different qubit operators at the same working point. Fig.~\ref{Fig_2} shows the balance between longitudinal modulation and co-rotating transverse driving for each propagation branch.
For each propagation direction, we quantify this balance with the parameter:
\begin{equation}\label{Eq:lambda_TL}
\Lambda_{\text{TL},\xi} = \frac{2}{\pi} \tan^{-1}\left(\frac{f_{L,\xi}}{f_{T,\xi}}\right) \,.
\end{equation}
The rates $f_{L,\xi}$ and $f_{T,\xi}$ are taken directly from the $g$-matrix modulation of Eq.~\eqref{Eq:gmatrix_fTL}.
The bounded quantity $\Lambda_{TL,\xi}$ visualizes the balance between longitudinal modulation and co-rotating transverse driving: $\Lambda_{TL,\xi}=0$ denotes $f_{L,\xi}=0$, whereas $\Lambda_{TL,\xi}=1$ denotes $f_{T,\xi}=0$ with $f_{L,\xi}\neq0$.
It does not encode the absolute coupling strength.

We consider two angular knobs.
The acoustic-axis angle $\theta_{\text{SAW}}$ resolves the normal and sagittal-shear strains of Eq.~\eqref{Eq:rayleigh_strain_schematic} into crystal-frame entries of $\delta\tilde{g}_\xi$ through Eqs.~\eqref{Eq:g_corrections} and~\eqref{Eq:rayleigh_to_crystal}.
The magnetic field angle $\theta_B$ changes the static Larmor axis $\hat{\bm{n}}$ onto which the resulting acoustic Larmor vector is projected.

The maps display an approximate fourfold diagonal texture.
In the present angular convention, the leading contribution is found to be proportional to $\cos[4(\theta_B+\theta_{\text{SAW}})+\varphi_0]$, consistent with the near invariance under a joint $90^\circ$ rotation of the magnetic field and the acoustic axis.
This structure reflects the cubic valence-band anisotropy and the different rotational characters of the normal- and shear-strain channels, while the realistic quasi-Rayleigh polarization and the finite-$k$ orbital averaging generate weaker harmonics.

The blue and red regions correspond to zeros of different projections of the acoustic Larmor vector.
Along a blue region, $f_{L,\xi}\simeq0$, so the SAW produces little qubit frequency modulation while retaining a finite transverse drive.
These are points of vanishing first-order longitudinal response to the modeled acoustic mode.
Along a red region, the co-rotating transverse projection cancels, $f_{T,\xi}\simeq0$, while the longitudinal response remains finite.
The acoustic and magnetic field orientations select between longitudinal modulation and co-rotating transverse driving.

\begin{figure*}[htbp!]
\centering
\includegraphics[width=\linewidth]{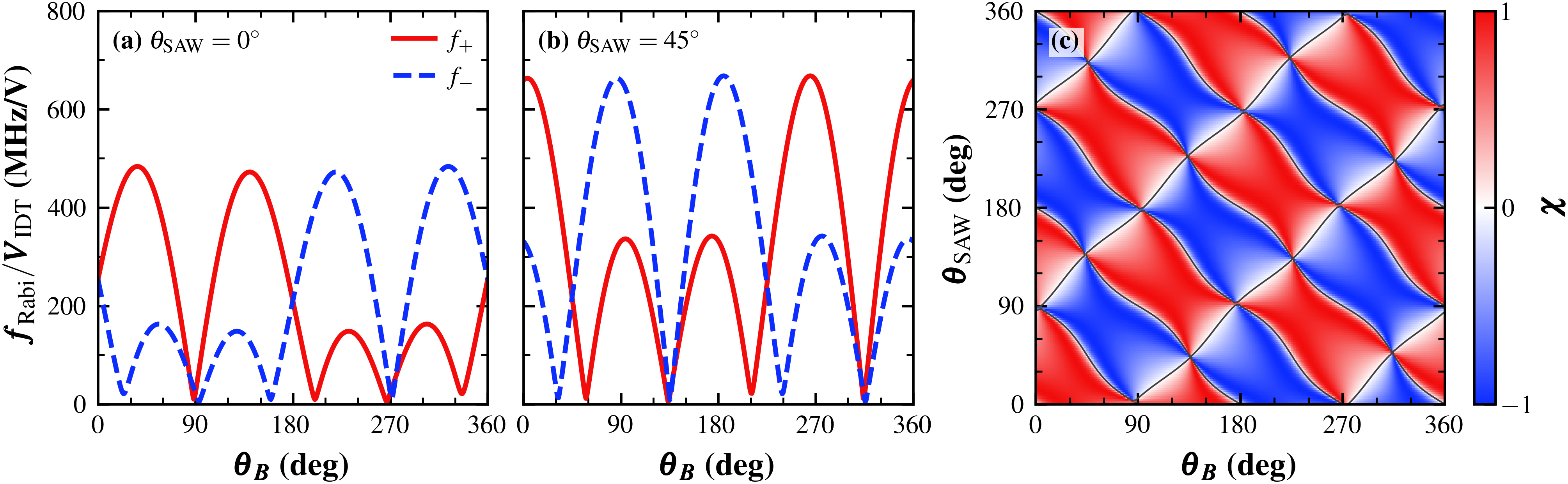}
\caption{FEM-derived directional transverse coupling and chirality. (a,b) Resonant Rabi frequencies $f_\pm\equiv f_{T,\xi=\pm1}$ per applied IDT voltage versus $\theta_B$ at $2$~GHz, using the directly calculated $\theta_{\text{SAW}}=0^\circ$ FEM field and its rotated $\theta_{\text{SAW}}=45^\circ$ counterpart, respectively.
(c) Chirality $\chi=(f_+^2-f_-^2)/(f_+^2+f_-^2)$ of Eq.~\eqref{Eq:chirality} as a function of $\theta_B$ and $\theta_{\text{SAW}}$.
Black contours denote $\chi=0$, that is, equal coupling of the two propagation directions, and not a vanishing absolute Rabi frequency.}
\label{Fig_3}
\end{figure*}
The nodal patterns in Fig.~\ref{Fig_2} are obtained from the complete complex FEM strain fields and their projections onto the qubit. Changing the launch direction changes the local strain amplitudes and phases, which can shift both the transverse and longitudinal nodal lines. The ideal Rayleigh model provides a qualitative interpretation of the dominant strain channels.

The magnetic field determines the handedness of the Larmor precession, giving the reciprocity relation $\Lambda_{\text{TL},+}(\bm{B})=\Lambda_{\text{TL},-}(-\bm{B})$ rather than equality of the two branches at fixed $\bm{B}$.
We find the contrast $\Lambda_\text{TL,+}-\Lambda_\text{TL,-}$ to range between $-0.63$ and $+0.63$, indicating there are magnetic field orientations where one incoming branch has $f_{L,\xi}\simeq0$, while the counterpropagating branch is dominated by longitudinal modulation, and vice versa.
However, there is no configuration where $\Lambda_\text{TL,+}-\Lambda_\text{TL,-}=\pm1$, which would indicate purely transverse and longitudinal response to the respective counterpropagating waves.

\subsection{Directional Rabi driving and acoustic chirality}
\label{Sec_3_2_directional_saw}
The limiting values of $\Lambda_{\text{TL},\xi}$ specify operator composition. The operator-composition maps should be interpreted together with the absolute transverse rates.
Fig.~\ref{Fig_3} shows that the two counter-propagating FEM-driven fields can drive the same Zeeman transition at very different rates, including working points where one field has a near-zero transverse Rabi rate.

The branch-dependent displacement of the transverse nodes already visible in Fig.~\ref{Fig_2} is the origin of this directional contrast.
In the full FEM calculation, changing the launch direction changes the relative amplitudes and phases with which the normal- and shear-strain channels project onto the co-rotating transverse Larmor component.
Their interference can therefore be constructive for one launch direction and destructive for the other.
The magnetic field fixes the handedness of the Larmor precession and selects which driven acoustic field is better matched to the transition.
The ideal Rayleigh parity relations provide a qualitative guide to this interference.

For $\theta_{\text{SAW}}=0^\circ$, each propagation branch exhibits four deep minima as the in-plane magnetic field direction is varied.
The two branches satisfy $f_{T,-}(\theta_B,0^\circ)=f_{T,+}(-\theta_B,0^\circ)$.
Rotating the acoustic axis to $45^\circ$ reorganizes these minima without changing their number.
The strong suppression for a magnetic field perpendicular to the propagation direction follows from the ideal Rayleigh limit, in which the acoustic modulation is purely longitudinal at this working point.
Device-induced transverse motion, nonuniform strain, and additional complex shear components weakly break the ideal symmetry and lift the node in the FEM result.
The residual minima are therefore remnants of a symmetry-protected Rayleigh node rather than accidental cancellations.

The directional contrast in Fig.~\ref{Fig_3}(c) is characterized by the chirality $\chi$ of Eq.~\eqref{Eq:chirality}.
Because $\chi$ is normalized, a large $|\chi|$ can result from suppressing one branch rather than enhancing the other.
A useful chiral working point requires both a large directional contrast and a finite surviving transverse rate, so panel (c) must be read together with panels (a) and (b).

A numerical Fourier decomposition of the unnormalized rate difference $f_{T,+}^2-f_{T,-}^2$ reveals the global angular structure. Its dominant contribution is proportional to $\sin(3\theta_{\text{SAW}}+\theta_B)+\sin(\theta_{\text{SAW}}+3\theta_B)$. This first-third-harmonic interference generates four principal symmetry-related zero directions and the diagonal lobe structure. Weaker Fourier components distort this ideal pattern, curving and locally splitting the zero contours. The pattern is consistent with the angular winding of the $J=3/2$ hole states and the coherent interference of distinct normal- and shear-strain Bir-Pikus channels in a cubic crystal.
Red and blue identify which direction couples more strongly; the black contours indicate equal directional coupling, not necessarily a vanishing absolute Rabi rate.

In the two-IDT geometry of Fig.~\ref{Fig_1}(a), the same qubit can be made bright to the SAW launched from one side and nearly dark to the reciprocal wave launched from the other.
In the quantized description developed in Sec.~\ref{Sec_4_spin_phonon}, the same transverse contrast becomes directional absorption and emission.
\begin{figure*}[htbp!]
\centering
\includegraphics[width=\linewidth]{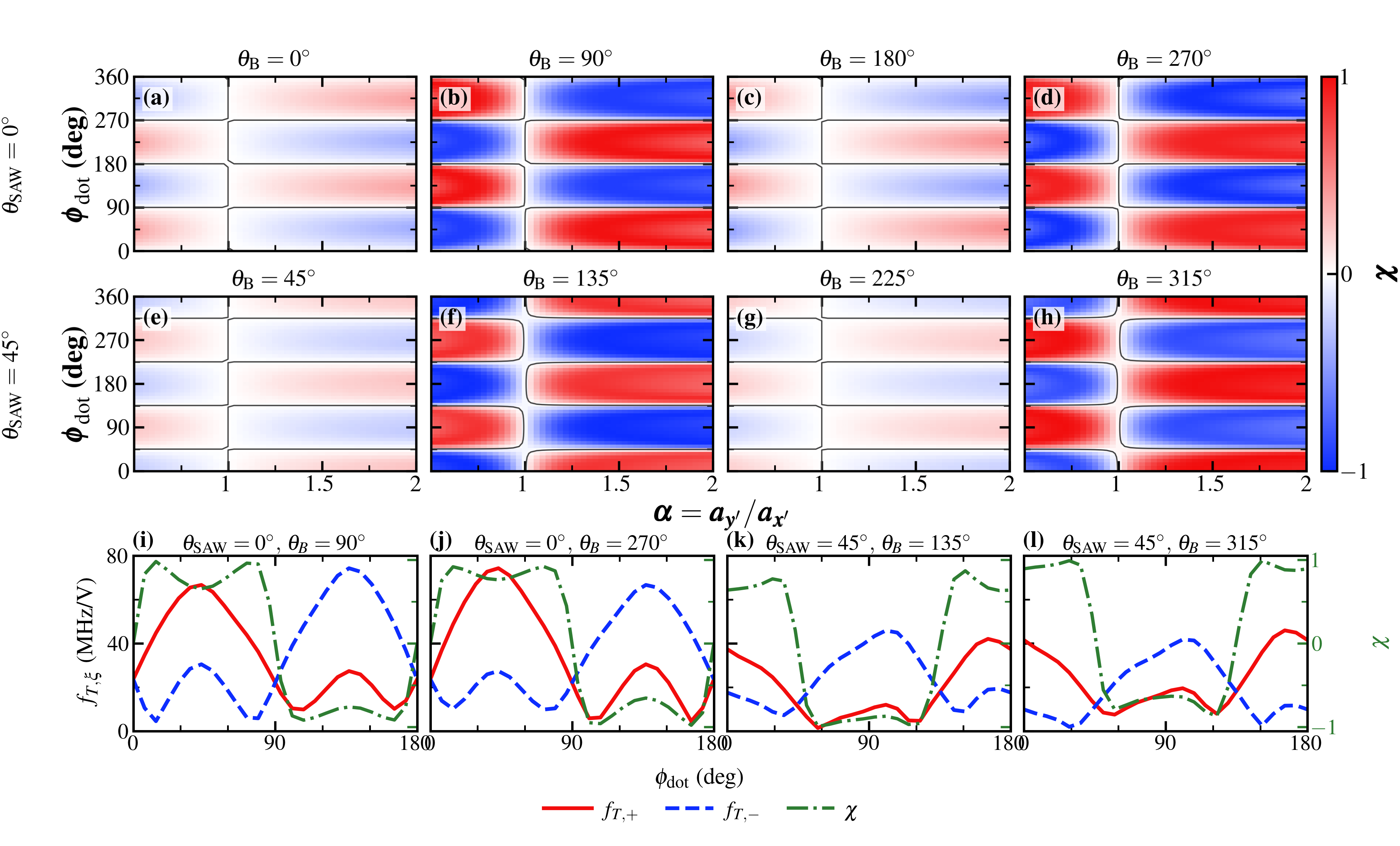}
\caption{FEM-calculated chirality and direction-resolved transverse coupling under electrostatic reshaping of the dot.
(a-d) $\chi$ as a function of the fixed-area aspect ratio $\alpha = a_{y'}/a_{x'}$ and dot orientation $\phi_{\text{dot}}$, for $\theta_{\text{SAW}} = 0^\circ$ and $\theta_B = 0^\circ$, $90^\circ$, $180^\circ$, and $270^\circ$, respectively.
(e-h) Corresponding maps for $\theta_{\text{SAW}} = 45^\circ$ and $\theta_B = 45^\circ$, $135^\circ$, $225^\circ$, and $315^\circ$.
Red and blue denote positive and negative chirality, respectively, and the grey contours indicate $\chi = 0$.
(i-l) One-dimensional cuts at $\alpha = 2$ through panels (b), (d), (f), and (h), respectively, showing $f_{T,+}$ (red solid), $f_{T,-}$ (blue dashed), and $\chi$ (green dash-dotted) as $\phi_{\text{dot}}$ varies over one non-redundant $180^\circ$ period.
The transverse rates are quoted per volt of differential peak IDT drive.}
\label{Fig_4}
\end{figure*}

\subsection{Electrical control through quantum dot confinement}
\label{Sec_3_3_confinement}
The preceding angular maps use a circular dot.
Local gates provide an additional control because changing the confinement modifies both the static $g$-matrix $\tilde{g}_0$, and hence the Larmor axis, and the acoustic correction $\delta\tilde{g}_\xi$. We find that, over the range considered, varying the vertical electric field has little effect on the acoustic response. 
We examine in Fig.~\ref{Fig_4} whether the preferred propagation direction can be reversed by reshaping the dot while keeping the physical SAW mode fixed, and whether this reversal preserves a finite transverse coupling.

The confinement is parameterized at fixed area by $\alpha = a_{y'}/a_{x'}$, $a_{x'} = a_0/\sqrt{\alpha}$, $a_{y'} = a_0\sqrt{\alpha}$, with $\alpha\in[0.5,2]$, and by the angle $\phi_{\text{dot}}$ between the principal axis of the ellipse and the crystal axes.
Here and below, unprimed axes $x,y$ denote the crystal frame used for the $g$-matrix entries and Bir-Pikus strain components of Sec.~\ref{Sec_2_model}.
Primed axes $x',y'$ denote the principal axes of the dot, rotated from the crystal frame by $\phi_{\text{dot}}$.
The FEM strain field and the device coordinates are kept fixed, while the static and acoustic qubit quantities are recalculated at every $(\alpha,\phi_{\text{dot}})$ point as described above.

For a strongly confined heavy-hole doublet, the leading in-plane entries of the static $g$-matrix have the form~\cite{Michal2021,Abadillo2023}:
\begin{subequations}
\label{Eq:static_g_confinement}
\begin{align}
\tilde{g}_{0,xx} &= 3q + \frac{6\left(\lambda\expval**{p_x^2}-\lambda'\expval**{p_y^2}\right)}{m_0\left(E_{\text{LH}}-E_{\text{HH}}\right)} \,,\\
\tilde{g}_{0,yy} &= -3q - \frac{6\left(\lambda\expval**{p_y^2}-\lambda'\expval**{p_x^2}\right)}{m_0\left(E_{\text{LH}}-E_{\text{HH}}\right)} \,,\\
\tilde{g}_{0,xy} &= -\tilde{g}_{0,yx} = -\frac{12\gamma_3\kappa\expval**{p_xp_y}}{m_0\left(E_{\text{LH}}-E_{\text{HH}}\right)} \,.
\end{align}
\end{subequations}
Here $q$ is the cubic Zeeman parameter, and $\lambda$ and $\lambda'$ are material-dependent combinations of Luttinger and Zeeman parameters.
The parameter $\gamma_3$ is a Luttinger parameter entering the static $g$-matrix, and $\kappa$ is the valence-band Zeeman parameter of Eq.~\eqref{Eq:g_corrections}.

In the principal frame of the dot the harmonic ground state gives $\expval**{p_{x'}^2}=\hbar^2/(2a_{x'}^2)$ and $\expval**{p_{y'}^2}=\hbar^2/(2a_{y'}^2)$, with $\expval**{p_{x'}p_{y'}}=0$.
After these averages are rotated into the crystal frame, the fixed-area deformation of the dot gives $\expval**{p_x^2}-\expval**{p_y^2} = \zeta\cos(2\phi_{\text{dot}})$, and $\expval**{p_xp_y} = \zeta\sin(2\phi_{\text{dot}})/2$, with the momentum anisotropy $\zeta = \expval**{p_{x'}^2}-\expval**{p_{y'}^2} = (\hbar^2/2a_0^2)(\alpha-\alpha^{-1})$.
Elongating the dot changes the in-plane entries of the static $g$-matrix.
Rotating it transfers weight from the diagonal to the antisymmetric entries and rotates the Larmor axis $\hat{\bm{n}}$ away from a generic magnetic field direction.

At $\alpha=1$, the in-plane orientation of the harmonic confinement is physically irrelevant and $\zeta=0$.
The orientation-dependent diagonal anisotropy and the antisymmetric off-diagonal correction disappear, while an orientation-independent diagonal confinement correction remains, giving the nearly orientation-independent central line of Fig.~\ref{Fig_4}.
Moving away from $\alpha=1$ splits the in-plane momentum variances, changes the static $g$-matrix and the Larmor axis, and modifies the acoustic $g$-matrix response sampled by the same FEM field.

When $\alpha \neq 1$, the confinement-induced off-diagonal momentum moment is defined in the crystal frame and obeys $\expval**{p_x p_y}=\zeta \sin(2\phi_{\text{dot}})/2$.
It vanishes when a principal dot axis is aligned with a crystal axis, independently of $\theta_{\text{SAW}}$.
For $\theta_{\text{SAW}}=0^\circ$, the crystal and acoustic axes coincide, so these zeros are consistent with the principal zero contours in the upper row of Fig.~\ref{Fig_4}.
For nonzero $\theta_{\text{SAW}}$, however, the zero contours of $\chi$ are determined by the complete direction-resolved transverse matrix elements.
Rotating the dot changes both the static Larmor axis and the projection of the FEM-derived acoustic $g$-matrix, while the normal- and shear-strain channels interfere with angle-dependent amplitudes and phases.
The shifted contours and alternating sign in the lower row therefore arise from the full LKBP-FEM response.
Reversing the magnetic field approximately exchanges the preferred propagation direction, producing the nearly sign-reversed patterns of Figs.~\ref{Fig_4}(a), (c), and (e), (g).

The largest $|\chi|$ values occur near the Rayleigh transverse-coupling nodes, that is, for the near-perpendicular field configurations of Figs.~\ref{Fig_4}(b) and (d) for $\theta_{\text{SAW}}=0^\circ$, and of Figs.~\ref{Fig_4}(f) and (h) for $\theta_{\text{SAW}}=45^\circ$.
The $\alpha=2$ cuts in Figs.~\ref{Fig_4}(i–l) show the corresponding transverse rates.
Across the four configurations, rotating the dot exchanges the stronger of $f_{T,+}$ and $f_{T,-}$.
At the extrema of $|\chi|$, the brighter branch retains a coupling of $23$–$48$ MHz/V, whereas the weaker branch is reduced to $1.4$–$9.9$ MHz/V.
The sign reversals in the chirality maps represent a gate-controlled exchange of the bright propagation direction.

These results establish the gate-tunable principle of the device.
The physical SAW polarization and acoustic axis are fixed by the selected IDT or acoustic mode, while the magnetic field orientation provides a global Larmor reference for the array.
Local gates then reshape each dot and modify both its static and its acoustic $g$-matrix response.
The same reciprocal SAW pair can consequently act as a predominantly transverse drive, a longitudinal modulator, a left-selective drive, or a right-selective drive at different gate-defined working points.

\section{Spin-phonon interaction}
\label{Sec_4_spin_phonon}
We now quantize the acoustic field to connect the classical response of Secs.~\ref{Sec_2_model} and~\ref{Sec_3_results} to spin-phonon coupling.
The same strain-induced $g$-matrix modulation applies to traveling and cavity-confined modes and retains the electrical and magnetic programmability established above.
Classically, the two quadratures $\delta\tilde{g}^{(c,s)}_\xi\bm{B}$ trace an acoustic Larmor ellipse around the static Larmor axis $\hat{\bm{n}}$.
Throughout this section, the quantized displacement, strain, and Larmor-vector fields are operators.
The displacement field of the traveling mode $(\xi,k)$ is expanded in bosonic operators as $\bm{u}(\bm{r})=\bm{u}_{\xi k}(\bm{r})\hat{a}_{\xi k}+\bm{u}^{*}_{\xi k}(\bm{r})\hat{a}^{\dagger}_{\xi k}$.
Here $\hat{a}^{\dagger}_{\xi k}$ ($\hat{a}_{\xi k}$) creates (annihilates) one phonon of frequency $\omega_{\xi k}$ in that mode.
The mode profile $\bm{u}_{\xi k}(\bm{r})$ is normalized to the zero-point energy $\hbar\omega_{\xi k}/2$~\cite{Li2020HolePhonon}.

Each strain component at the dot becomes an operator, $\epsilon_{ij}=\epsilon^{\text{zpf}}_{ij,\xi k}\hat{a}_{\xi k}+\epsilon^{\text{zpf}*}_{ij,\xi k}\hat{a}^{\dagger}_{\xi k}$.
Its complex zero-point amplitude $\epsilon^{\text{zpf}}_{ij,\xi k}$ is the strain of the energy-normalized mode.
The phases of these amplitudes retain the mutual phases among the strain components fixed classically by the elastic eigenmode.
These strain components fluctuate even when the mode is in the phonon vacuum.
Since the $g$-matrix corrections of Eq.~\eqref{Eq:g_corrections} are linear in the strain components, the $g$-matrix modulation inherits this operator structure.
For a normalized traveling mode $(\xi,k)$, we denote the complex zero-point $g$-matrix amplitude by $\delta\tilde{g}^{\text{zpf}}_{\xi k}$.
It is obtained by energy-normalizing the same FEM mode used for the classical calculation.
The associated zero-point Larmor-vector amplitude is then given by $\delta\bm{\Omega}^{\text{zpf}}_{\xi k}=(\mu_{\text{B}}/\hbar)\delta\tilde{g}^{\text{zpf}}_{\xi k}\bm{B}$.
In terms of the phonon operators, the quantized counterpart of Eq.~\eqref{Eq:gmatrix_quadratures} is:
\begin{equation}
\delta\bm{\Omega}_{\xi k} = \delta\bm{\Omega}^{\text{zpf}}_{\xi k}\hat{a}_{\xi k} + \delta\bm{\Omega}^{\text{zpf}*}_{\xi k}\hat{a}^\dagger_{\xi k} \,.
\end{equation}

A phonon excitation can influence the three geometric components of this same Larmor vector.
Its transverse part decomposes into two circular components of opposite handedness around $\hat{\bm{n}}$.
The component matched to the qubit precession defines the co-rotating coupling $g^{\text{co}}_{\xi k}$, whereas the opposite-handed component defines the counter-rotating coupling $g^{\text{ctr}}_{\xi k}$.
Its projection along the static Larmor axis $g^{\parallel}_{\xi k} = \hat{\bm{n}}\cdot\delta\bm{\Omega}^{\text{zpf}}_{\xi k}/2$ produces a phonon-dependent modulation of the qubit splitting.
The one-mode Hamiltonian is:
\begin{equation}\label{Eq:generalized_rabi}
\begin{aligned}
\frac{H_{\xi k}}{\hbar} = &\frac{\omega_q}{2}\sigma_z + \omega_{\xi k}\hat{a}^\dagger_{\xi k}\hat{a}_{\xi k} \\
& + \left[\left(g^{\text{co}}_{\xi k}\sigma_+ + g^{\text{ctr}}_{\xi k}\sigma_- + g^{\parallel}_{\xi k}\sigma_z\right)\hat{a}_{\xi k} + \text{H.c.}\right] \,,  
\end{aligned}
\end{equation}
a generalized anisotropic Rabi interaction, where anisotropic refers to unequal rotating and counter-rotating coupling strengths.
A circular transverse Larmor field gives the Jaynes-Cummings (JC) or anti-JC limit depending on its handedness.
A linear transverse field gives equal rotating and counter-rotating couplings and hence the ordinary Rabi model.
A generic tilted ellipse gives unequal transverse couplings together with a longitudinal term.

We can connect this quantum limit with the classical SAW limit to understand the underlying physics.
For a coherent state of amplitude $\alpha_{\xi k}$, the three zero-point couplings reproduce the classical responses of Sec.~\ref{Sec_2_model} through the following relations:
\begin{subequations}
\begin{align}
2\pi f_{T,\xi}=&2|g^{\text{co}}_{\xi k}\alpha_{\xi k}|, \\
2\pi f_{\text{ctr},\xi}=&2|g^{\text{ctr}}_{\xi k}\alpha_{\xi k}|, \\
2\pi f_{L,\xi}=&4|g^{\parallel}_{\xi k}\alpha_{\xi k}|.
\end{align}
\end{subequations}
Here $f_{\text{ctr},\xi}$ is the counter-rotating transverse scale rather than an additional resonant Rabi frequency.
Resonant driving measures $f_{T,\xi}$, whereas $f_{\text{ctr},\xi}$ describes the opposite-handed component retained by the exact Hamiltonian and discarded by the rotating-wave approximation.
Appendix~\ref{App:RayleighMode} gives the matrix-element definitions used to evaluate the classical transverse and longitudinal rates.

The quantum couplings are controlled by the same knobs as the classical response, and are not independent phenomenological parameters. Dot reshaping jointly tunes $\hat{\bm{n}}$ through the static $g$-matrix and $\delta\tilde{g}^{\text{zpf}}_{\xi k}$ through the acoustic response.
Figs~\ref{Fig_2} and~\ref{Fig_3} characterize the longitudinal/co-rotating balance and the direction-resolved co-rotating strengths of the FEM-driven fields, respectively.
Fig~\ref{Fig_4} shows that dot-shape gates can reverse this directional imbalance.

For a pair of equally normalized traveling modes related by time reversal, the counter-rotating strength of one direction equals the co-rotating strength of the other at the same magnetic field: $|g^{\text{ctr}}_{+,k}(\bm{B})|=|g^{\text{co}}_{-,k}(\bm{B})|$.
We therefore define the transverse anisotropy of a selected traveling mode as:
\begin{equation}
\mathcal{A}_{\xi} = \frac{|g^{\text{co}}_{\xi k}|^2-|g^{\text{ctr}}_{\xi k}|^2}{|g^{\text{co}}_{\xi k}|^2+|g^{\text{ctr}}_{\xi k}|^2} \,.
\end{equation}
Under this reciprocal-mode condition, this expression reduces to $\mathcal{A}_\pm=\pm\chi$, where $\chi$ is the chirality defined in Eq.~\eqref{Eq:chirality}.

In the ideal reciprocal traveling-mode limit, the chirality maps can also be interpreted as maps of the transverse Rabi anisotropy. Regions where $\chi\to+1$ place the transverse sector of the $+$-propagating mode near the JC limit ($\mathcal{A}_+\to+1$) and its reciprocal partner near the anti-JC limit ($\mathcal{A}_-\to-1$).
A reversal of $\chi$ in Fig.~\ref{Fig_4} exchanges these roles between the two directions.
$\chi=0$ contours mark equal co- and counter-rotating transverse strengths, corresponding to linear transverse polarization and the ordinary Rabi limit of the transverse sector.
The longitudinal coupling may remain finite.
A gate-voltage or field-angle trajectory across these features tunes the transverse sector of a selected traveling mode continuously from JC through Rabi to anti-JC, with the longitudinal coupling $g^{\parallel}_{\xi k}$ evolving along the same trajectory.
This interpretation applies to a direction-resolved traveling mode; a nondegenerate standing cavity mode can combine the two directions and recover equal transverse sectors.

For the driven FEM fields, $\chi$ quantifies the directional contrast of the co-rotating coupling.
Since these opposite-launch solutions need not to form a pointwise time-reversal pair, no equality of their longitudinal amplitudes is imposed.
The exact reciprocal-mode relation $A_{\pm}= \pm\chi$ applies to the equally normalized traveling-mode pair defined above, whereas the precise transverse anisotropy of the FEM fields requires evaluating the co- and counter-rotating couplings separately.

The generalized interaction in Eq.~\eqref{Eq:generalized_rabi} is valid for any spin-phonon coupling strength and applies equally to present devices and to future architectures with enhanced phonon confinement.
The complete Hamiltonian identifies the quantum interaction associated with each component of the gate-tunable acoustic Larmor ellipse and establishes the correspondence between classical chirality and the quantum anisotropic Rabi model.

Our device-specific hybrid estimate combines the FEM-LKBP open-SAW projection with the PnC cavity model of Appendix~\ref{App:BellStateDynamics}.
It places the PnC architecture of Fig.~\ref{Fig_1}(b) in the weak-coupling regime, with a local defect-mode coupling scale of order $1$~kHz.
More generally, recent estimates for Ge hole-spin devices place single-phonon coupling strengths at or below the ${\sim}10~\mathrm{MHz}$ scale~\cite{Mei2025,Myronov2026}, well below the GHz phonon frequency, so counter-rotating terms are negligible for the near-resonant protocols considered below.
The rotating-wave approximation then reduces Eq.~\eqref{Eq:generalized_rabi} to:
\begin{equation}
\label{Eq:quantized_acoustic_vertex}
H_{\mathrm{int}}^{\mathrm{JC}} = \sum_{\xi=\pm1}\sum_{k>0}\left(\hbar g_{\xi k}^{\mathrm{co}}\sigma_{+}\hat{a}_{\xi k}+\text{H.c.}\right) \,.
\end{equation}

For a discrete cavity, Eq.~\eqref{Eq:quantized_acoustic_vertex} reduces to the corresponding selected mode, whereas for a direction-resolved itinerant continuum it gives spontaneous emission into the two propagation branches.
In the latter case, resonant emission probes the co-rotating sector alone.
For an equally normalized reciprocal pair, the directional phonon densities of states are equal and $\Gamma_{\xi}=2\pi |g_{\xi k_q}^{\mathrm{co}}|^{2}\rho_{\xi}(\omega_q)$.
Here $k_q$ is defined by $\omega_{\xi k_q}=\omega_q$, and $\rho_\xi$ is the density of states of branch $\xi$ per unit angular frequency.
Consequently, the one-phonon emission contrast is:
\begin{equation}
\chi^{\mathrm{em}}=\frac{\Gamma_{+}-\Gamma_{-}}{\Gamma_{+}+\Gamma_{-}} \,.
\end{equation}
This contrast equals the classical chirality of Eq.~\eqref{Eq:chirality}.
The directional zeros and sign reversals identified in Sec.~\ref{Sec_3_results} survive in the quantum limit.
Section~\ref{Sec_5_2_bell} uses the weakly coupled PnC defect mode as a monitored emission channel for heralded entangled-state initialization.

\section{Applications}
\label{Sec_5_applications}
We now use the common acoustic $g$-matrix description in two applications: bichromatic coherence protection with classical SAWs and heralded Bell-state initialization through a quantized PnC mode.

\subsection{Chirality-assisted coherence protection}
\label{Sec_5_1_coherence}
We designate the $\xi=+1$ transverse branch as bright.

A primary tone at frequency $f_1\simeq f_q$ is launched through the bright branch and drives the qubit transversely with Rabi frequency $f_{R1,+}$.
A second, much lower-frequency tone at $f_2$ is launched from the opposite side and predominantly modulates the qubit splitting with peak longitudinal amplitude $f_{L2,-}$.
When $f_2\simeq f_{R1,+}$, this longitudinal modulation resonantly couples the states created by the primary drive and opens a second-dressed gap.
The resulting two-frequency drive may be viewed as a bichromatic Floquet protocol~\cite{cai2012robust,Bosco2023}.
The primary dressing converts slow fluctuations of the bare qubit splitting into noise transverse to the first dressed-state axis.
It suppresses their leading dephasing contribution, but fluctuations of the primary Rabi amplitude still change the first dressed splitting to first order.

The auxiliary tone suppresses this remaining first-order sensitivity: in the first-dressed basis, a modulation that was longitudinal for the bare qubit is transverse and can dress the qubit a second time.
At the second resonance, fluctuations of the primary Rabi frequency are also transverse and enter the final splitting only to second order.
The auxiliary-tone amplitude noise then becomes the leading first-order control-noise channel.

In the static-qubit basis, $\sigma_z$ is aligned with the static Larmor vector $\bm{\Omega}_0$ given by the magnetic field orientation and the hole $g$-matrix.
A SAW driving field with frequency $f_1$ is sent through one of the IDTs in Fig.~\ref{Fig_1}(a).
We choose the phase of the primary tone so that its co-rotating component points along $\sigma_x$.
We transform to the rotating frame with the unitary $U_1(t)=\mathrm{e}^{-\mathrm{i}\pi f_1 t\sigma_z}$:
\begin{equation}\label{Eq:bichromatic_first_frame}
\begin{aligned}
\frac{H_1(t)}{h} =& \frac{\Delta_1+\delta\Delta_1(t)}{2}\sigma_z + \frac{f_{R1,+}+\delta f_{R1}(t)}{2}\sigma_x\\
&+ \frac{f_{L2,-}[1+a_2(t)]\cos(2\pi f_2t+\varphi_2)}{2}\sigma_z  \,.
\end{aligned}
\end{equation}
Here $\Delta_1=f_q-f_1$, and $f_{R1,+}$ is the Rabi frequency of the incoming SAW transverse mode.
The primary-drive amplitude fluctuation in this frame is expressed as $\delta f_{R1}(t)=f_{R1,+}a_1(t)$.
The instantaneous detuning noise between the qubit and the primary tone is denoted by $\delta\Delta_1(t)$.
It contains the bare qubit-frequency noise together with the primary source-frequency noise and acoustic-path fluctuations.
In this frame, we include the auxiliary longitudinal incoming SAW with qubit frequency driving $f_{L2,-}$, frequency $f_2$, and fractional amplitude noise denoted by $a_2(t)$.

At exact resonance, $\Delta_1=0$, we can redefine the qubit axes by a Hadamard rotation $R_H=\mathrm{e}^{\mathrm{i}\pi(\sigma_x+\sigma_z)/(2\sqrt{2})}$, defining the first-dressed quantization axis.
The first-dressed Hamiltonian then has the simple physical form:
\begin{equation}
\begin{aligned}
\frac{H_{\text{d1}}(t)}{h} = & \frac{f_{R1,+}+\delta f_{R1}(t)}{2}\sigma_z +\frac{\delta\Delta_1(t)}{2}\sigma_x \\
&+ \frac{f_{L2,-}[1+a_2(t)]\cos(2\pi f_2t+\varphi_2)}{2}\sigma_x \,.  
\end{aligned}
\end{equation}
This expression makes the protection of the first tone explicit: the bare-qubit detuning noise, which was longitudinal in the static-qubit basis, is transverse to the first-dressed axis.
It also shows why the auxiliary longitudinal modulation can drive transitions within the first-dressed doublet.

The second rotating frame follows that doublet at frequency $f_2$.
Applying $U_2(t)=\mathrm{e}^{-\mathrm{i}\pi f_2t\sigma_z}$ to $H_{\text{d1}}$ and making a rotating-wave approximation to the auxiliary drive gives:
\begin{equation}
\label{Eq:bichromatic_second_frame}
\frac{H_2(t)}{h} = \frac{\Delta_2+\delta\Delta_2(t)}{2}\sigma_z+\frac{f_{R2}+\delta f_{R2}(t)}{2}\sigma_x \,.
\end{equation}
The second-frame detuning is $\Delta_2=f_{R1,+}-f_2$, while $f_{R2}=f_{L2,-}/2$, and $\delta f_{R2}(t)=f_{R2}a_2(t)$.
The second-frame detuning noise is $\delta\Delta_2(t)=\delta\nu_1(t)-\delta f_{2,\text{rel}}(t)$, where $\delta\nu_1(t)$ is the fluctuation of the first-dressed splitting and $\delta f_{2,\text{rel}}(t)$ contains the relative source-frequency noise and differential acoustic-path fluctuations of the two tones.

The two resonance conditions are $f_1=f_q$ and $f_2=f_{R1,+}$.
At these conditions, the instantaneous first- and second-dressed splittings are:
\begin{subequations}
\begin{align}
\nu_1(t)&=\sqrt{[f_{R1,+}+\delta f_{R1}(t)]^2+\delta\Delta_1^2(t)} \,,\\
\nu_2(t)&=\sqrt{[f_{R2}+\delta f_{R2}(t)]^2+\delta\Delta_2^2(t)} \,.
\end{align}
\end{subequations}
Expanding them to second order and subtracting their mean shifts gives:
\begin{subequations}\label{Eq:bichromatic_noise_hierarchy}
\begin{align}
\delta\nu_1(t) &= \delta f_{R1}(t)+\frac{\delta\Delta_1^2(t)-\expval**{\delta\Delta_1^2}}{2f_{R1,+}} \,,\\
\delta\nu_2(t) &= \delta f_{R2}(t)+\frac{\delta\Delta_2^2(t)-\expval**{\delta\Delta_2^2}}{2f_{R2}} \,.
\end{align}
\end{subequations}
The first dressing suppresses bare-qubit detuning noise to second order but leaves primary amplitude noise at first order.
The second dressing converts this residual first-dressed detuning, including the primary amplitude noise, into a second-order contribution.
Its useful strength is ultimately limited by auxiliary amplitude noise, residual quadratic noise, and dressed-state relaxation.
Appendix~\ref{App:BichromaticCoherence} gives the spectra, relaxation rates, and numerical procedure used below.

Using the effective splitting fluctuations in Eqs.~\eqref{Eq:bichromatic_noise_hierarchy}, we construct the one-sided effective longitudinal frequency-noise spectrum $S_{\nu,m}(f)$ in each frame $m=0,1,2$, corresponding to the bare, first-dressed, and second-dressed qubits.
The normalized coherence envelope is:
\begin{subequations}
\label{Eq:coherence}
\begin{align}
C_m(t) &= \exp\left[-K_m(t)-\frac{\Gamma_{1,m,\text{eff}}t}{2}\right] \,,\\
K_m(t) &= \frac{(2\pi)^2}{2}\int_{f_{\text{lo}}}^{f_{\text{hi}}}S_{\nu,m}(f)\left[\frac{\sin(\pi f t)}{\pi f}\right]^2 \dd{f} \,.
\end{align}
\end{subequations}
Here, $S_{\nu,m}(f)$ is the one-sided PSD of the longitudinal fluctuation of the bare ($m=0$), first-dressed ($m=1$), or second-dressed ($m=2$) splitting, and $K_m(t)$ is the corresponding filter-function dephasing exponent.
The effective relaxation rates are $\Gamma_{1,0,\text{eff}}=\Gamma_{\text{lab}}$, $\Gamma_{1,1,\text{eff}}=\Gamma_{\text{lab}}+\Gamma_{1\rho}$, and $\Gamma_{1,2,\text{eff}}=\Gamma_{\text{lab}}+\Gamma_{1\rho}+\Gamma_{1\rho\rho}$ for the bare, first-dressed, and second-dressed qubits, respectively.
Here $\Gamma_{\text{lab}}$ is the laboratory-frame relaxation rate, while $\Gamma_{1\rho}$ and $\Gamma_{1\rho\rho}$ are the first- and second-dressed relaxation rates, respectively.
The explicit PSD expressions for these dressed relaxation rates are given in Appendix~\ref{App:BichromaticCoherence}.

\begin{figure}[htbp!]
\centering
\includegraphics[width=\linewidth]{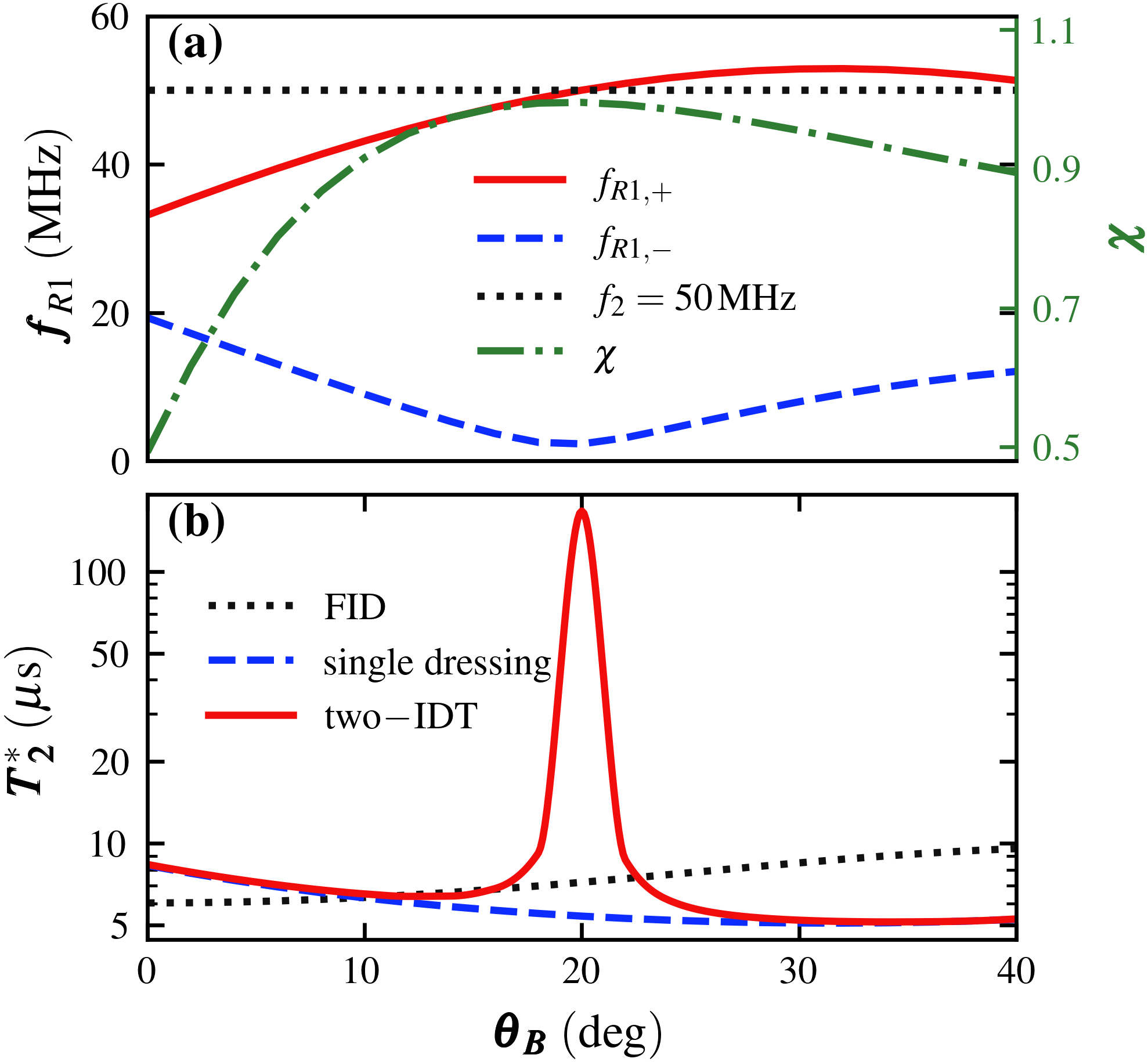}
\caption{Chirality-assisted bichromatic coherence protection.
(a) Primary Rabi frequencies $f_{R1,+}$ and $f_{R1,-}$ for the two reciprocal propagation branches, together with the chirality $\chi$, as functions of $\theta_B$.
The horizontal reference fixes $f_2=50$~MHz.
The acoustic controls are calibrated only at $\theta_B=20^\circ$, where the bright-branch splitting satisfies $f_{R1,+}=f_2$ while the reciprocal branch is nearly dark.
(b) Coherence times $T_{2,m}^*$ for bare free induction ($m=0$), single dressing ($m=1$), and bichromatic dressing ($m=2$), evaluated with the same noise and relaxation model.
Frequencies, powers, propagation directions, and acoustic geometry are held fixed throughout the angular scan.}
\label{Fig_5}
\end{figure}

We apply this protocol to the fixed-input angular scan shown in Fig.~\ref{Fig_5}.
We take $\alpha=1$, $f_1=2$~GHz, and $f_2=50$~MHz, with the matching condition $f_2=f_{R1,+}$ imposed only at the calibration angle.
The direction-resolved acoustic couplings are obtained from the FEM-LKBP calculation.
Both acoustic inputs are calibrated only at $\theta_B=20^\circ$.
The frequencies, external input settings, propagation directions, and acoustic geometry are then held fixed throughout the scan.
The electric field susceptibilities and noise model are specified in Appendix~\ref{App:BichromaticCoherence}.

Fig.~\ref{Fig_5}(a) shows how chirality creates the matching condition under fixed acoustic controls.
Near $\theta_B=20^\circ$, the normal- and shear-strain contributions interfere constructively for the $\xi=+1$ branch and destructively for the reciprocal branch.
The resulting chirality is $\chi=0.996$, corresponding to a $20$-fold contrast in Rabi frequency and a $470$-fold contrast in transition strength, while the bright branch retains a large absolute rate.
This last point is essential: a normalized chirality close to unity would not be useful if both branches were nearly dark.
At the selected angle, $f_{R1,-}\ll f_2$, so the fixed auxiliary tone is far detuned from the reciprocal-branch doublet.
Magnetic field rotation converts the microscopic interference responsible for chirality into selective activation of the second dressing.

The corresponding coherence times are shown in Fig.~\ref{Fig_5}(b).
At $\theta_B=20^\circ$, the bare free-induction value is $T_{2,0}^*=7.19~\mu\mathrm{s}$, whereas the single-dressed value is slightly shorter, $T_{2,1}^*=5.40~\mu\mathrm{s}$.
Although the primary tone suppresses bare-qubit detuning noise, its fractional amplitude noise remains longitudinal in the first-dressed basis and modulates $f_{R1,+}$ to first order.
The auxiliary tone removes this limitation only close to the second-frame resonance.
Away from it, the first-dressed detuning retains a longitudinal projection proportional to $\Delta_2^2/(\Delta_2^2+f_{R2}^2)$, whereas at $\Delta_2=0$ the primary-amplitude fluctuations are purely transverse to the second-dressed axis.
The remaining first-order channel is then dominated by auxiliary-tone amplitude noise, and the calculated coherence reaches $T_{2,2}^*=167~\mu\mathrm{s}$.

For extension to qubit arrays, the globally shared SAW could be pulse-engineered using the sinusoidally modulated protocol, improving robustness to site-to-site variations in qubit frequency and acoustic Rabi amplitude while local gate tuning preserves addressability~\cite{Hansen2021,Hansen2022}.

\subsection{Heralded Bell-state initialization in an asymmetric phononic cavity}
\label{Sec_5_2_bell}
Now, we focus on an example of application for the quantized chirality. We consider the device-anchored PnC estimate of Fig.~\ref{Fig_1}(b), whose kilohertz-scale coupling, while relatively weak, can be used as a heralded state-initialization primitive. Repeated trials initialize a phase-defined odd-parity two-spin state that can be used as the input to a subsequent measurement or locally controlled evolution.
For quasi-one-dimensional phononic guides with megahertz-scale spin-phonon coupling, Appendix~\ref{App:CascadeBenchmark} considers cascaded steady-state Bell stabilization.

We denote the lower and upper Zeeman eigenstates of each qubit by $\ket{0}$ and $\ket{1}$.
In $\ket{q_1q_2,n_c}$, the first and second entries label qubits 1 and 2, while $n_c$ is the cavity phonon number.
Two hole-spin qubits resonant with a common $2$~GHz defect mode are initialized in $\ket{11,0_c}$, and no coherent acoustic drive is applied during the heralding window.
Detection of the first phonon leaking through the shorter right mirror erases the which-qubit information and ideally prepares:
\begin{equation}
\ket{\psi_{\text{click}}}=\frac{g_1^*\ket{01}+g_2^*\ket{10}}{\sqrt{|g_1|^2+|g_2|^2}} \,,
\end{equation}
where $g_j$ is the local coupling of qubit $j$ to the defect mode introduced below.
In the near-resonant regime only the co-rotating sector contributes, so $g_j$ is effectively the co-rotating coupling $g^{\text{co}}_{\xi k}$ of Sec.~\ref{Sec_4_spin_phonon} for the counterpropagating wave components composing the confined mode, evaluated at the qubit position.

When $|g_1|=|g_2|$, the first registered phonon initializes the maximally entangled odd-parity Bell state as $\ket{\Psi_B(\phi_B)} = (\ket{01}+\mathrm{e}^{\mathrm{i}\phi_B}\ket{10})/\sqrt{2}$ where $\phi_B = -\arg(g_2/g_1)$.
The click-resolved model removes the timestamp-dependent dynamical phase with an ideal local $Z$ operation (Appendix~\ref{App:BellStateDynamics}).
Trajectories containing a second registered phonon during the feedback latency are discarded because the second emission produces $\ket{00}$.
After the feedback latency, the qubits are assumed to be detuned from the cavity.
The downstream receiver is modeled as an ideal time-resolved phonon detector; a microscopic phonon-to-electrical or phonon-to-microwave transduction chain is outside the present calculation.

Directional outcoupling is quantified by $\beta_{\text{R}}\equiv\kappa_{\text{R}}/\kappa$, with $\kappa=\omega_c/Q_{\text{L}}$. Here $\kappa_{\text{R}}$ is the cavity decay rate through the monitored right mirror.
The quantity $\beta_{\text{R}}$ is the fraction of the total cavity energy decay that leaves through that port, before detector inefficiency is included.
For a detector with efficiency $\eta_{\text{det}}$, the registered fraction is therefore $\beta_{\text{R}}\eta_{\text{det}}$.
The microscopic chirality $\chi$ of Eq.~\eqref{Eq:chirality} instead controls the relative coupling to the reciprocal $+k$ and $-k$ components of the defect mode and hence the spatial interference seen by the spin.
By contrast, $\beta_{\text{R}}$ is set by the mirror asymmetry and controls collection into the monitored output.
Chirality improves the balance of the two spin-cavity couplings, whereas the mirror asymmetry improves collection and heralding efficiency.

We use the fixed asymmetric PnC design described in Appendix~\ref{App:BellStateDynamics}, with $f_c=2$~GHz, $Q_{\text{L}}=4.0\times10^3$, and $\beta_{\text{R}}=0.95$.
The field-angle dependence and complex phases of the local cavity couplings are obtained from the direction-resolved co-rotating spin-phonon couplings calculated with the same FEM-LKBP model as the classical response.
Appendix~\ref{App:BellStateDynamics} combines the microscopic open-SAW projection with a two-dimensional cavity model and an imposed transverse focusing factor, giving the hybrid estimate $g_0/2\pi\sim1\,\mathrm{kHz}$ rather than a full three-dimensional defect-mode calculation.
For fixed cavity phase $\phi_c$ and qubit positions $x_j$, the local cavity couplings factorize as $g_j(\theta_B,x_j)=g_0(\theta_B)\mathcal{C}_j(\theta_B,x_j)$.
The dimensionless local interference factor $\mathcal{C}_j$ contains the normalized directional amplitudes, relative phases, and position dependence.
The common scale $g_0(\theta_B)$ carries the angle-dependent total strength of the coupling.
The same FEM-LKBP-derived couplings determine both factors, and no angle-dependent fitting is introduced.

The normalized local interference factor satisfies:
\begin{equation}
|\mathcal{C}(\theta_B,x)|^2=\frac{1}{2}\left[1+\sqrt{1-\chi^2(\theta_B)}\cos(2kx+\varphi)\right] \,,
\end{equation}
where $\varphi$ contains the relative phases of the direction-resolved coupling amplitudes and of the cavity mode.
For $\chi\simeq0$, the reciprocal contributions have comparable amplitudes and produce strong nodes and antinodes in the local coupling.
As $|\chi|\to1$, one contribution dominates the spin-cavity coupling and the spatial modulation of $|g_j|$ is suppressed, although the mechanical mode itself remains cavity confined.
Chirality therefore improves the tolerance of $|g_1|\simeq|g_2|$ to placement errors, rather than primarily increasing the heralding rate.

Before including decoherence, unmonitored loss, and feedback, the fidelity associated only with the coupling-amplitude imbalance is:
\begin{equation}
F_{\text{amp}}=\frac{1}{2}+\frac{|g_1g_2|}{|g_1|^2+|g_2|^2} \,.
\end{equation}
This quantity reaches unity for equal coupling magnitudes, and should be distinguished from the full conditional fidelity obtained from the click-resolved master equation of Appendix~\ref{App:BellStateDynamics}.
\begin{figure}[htbp!]
\centering
\includegraphics[width=\linewidth]{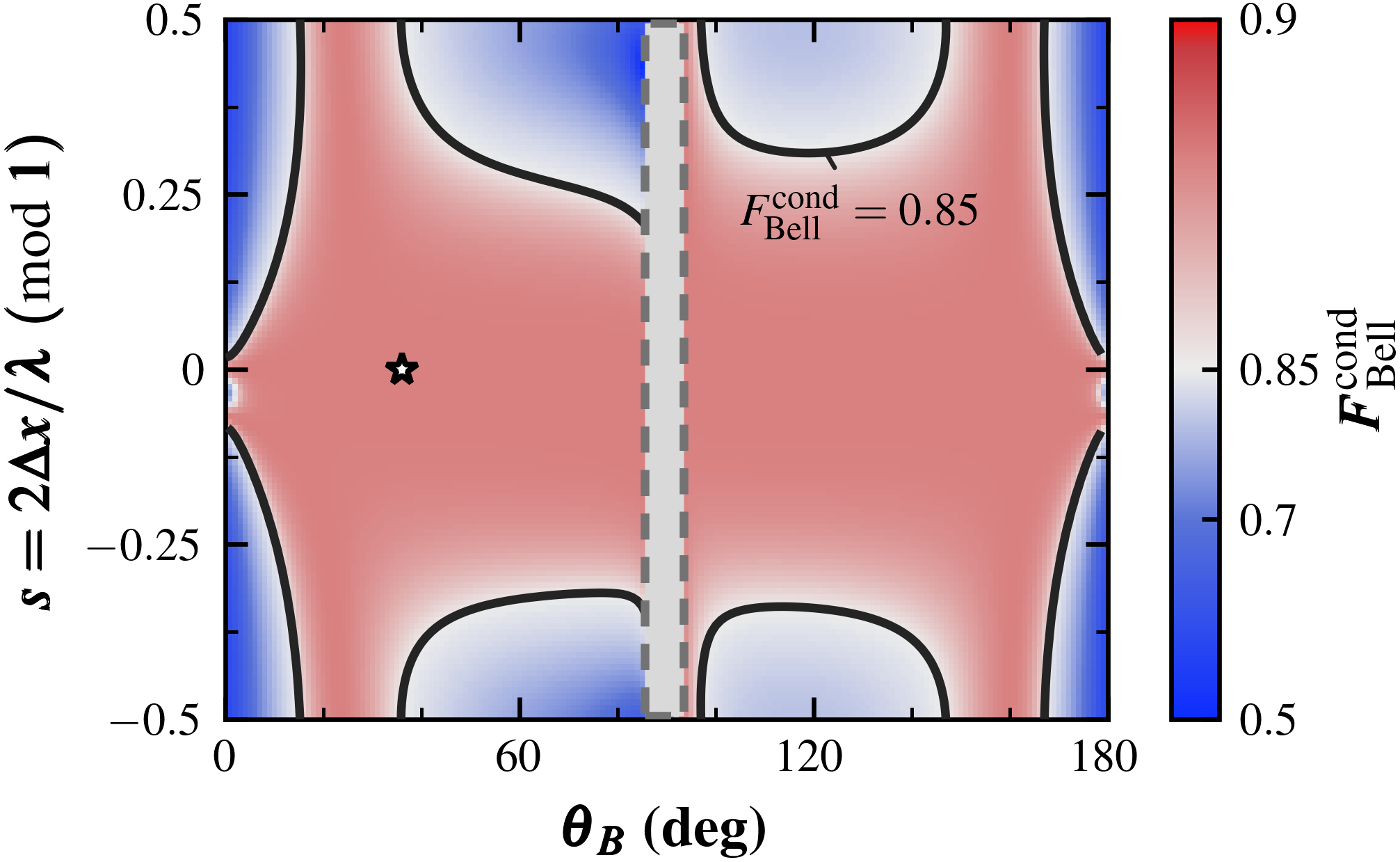}
\caption{Heralded Bell-state initialization in a shared $2$~GHz PnC defect mode.
Timestamp-corrected conditional Bell-state fidelity $F_{\text{Bell}}^{\text{cond}}$ as a function of magnetic-field angle $\theta_B$ and relative qubit displacement $s=2\Delta x/\lambda_{\text{PnC}}$.
The upper and lower boundaries represent the same placement phase.
The black contour marks $F_{\text{Bell}}^{\text{cond}}=0.85$, the grey band denotes the herald-dark angular interval, and the star marks the reference point $(\theta_B,s)=(36^\circ,0)$.}
\label{Fig_6}
\end{figure}

Figure~\ref{Fig_6} shows the timestamp-corrected conditional fidelity as a function of the magnetic-field angle and the relative placement phase $s=2(x_2-x_1)/\lambda_{\text{PnC}}$.
The coordinate is periodic because the standing-wave intensity repeats after $\lambda_{\text{PnC}}/2$, so the upper and lower boundaries represent the same placement phase.
At $s=0$, the two qubits sample equivalent points of the defect mode and satisfy $|g_1|=|g_2|$, producing the broad high-fidelity ridge.
Large $|\chi|$ broadens this ridge away from $s=0$ by suppressing the standing-wave contrast, whereas small $|\chi|$ produces stronger nodes, antinodes, and placement sensitivity, giving the low-fidelity lobes bounded by the black $F_{\text{Bell}}^{\text{cond}}=0.85$ contour.

Near $\theta_B=90^\circ$, the total FEM-derived transverse coupling is strongly suppressed, so the accepted heralding probability becomes negligible; this interval is shown as the grey herald-dark region.
At each $(\theta_B,s)$ point, the target phase is fixed by the microscopic ratio $g_2/g_1$ at that point; it is neither held fixed across the map nor optimized from the accepted density matrix.

At the reference point $(\theta_B,s)=(36^\circ,0)$, the calculation gives $F_{\text{Bell}}^{\text{cond}}\simeq0.88$ and an accepted heralding rate $R_{\text{acc}}\simeq82~\mathrm{s}^{-1}$.
Under the assumed unit detection efficiency, zero dark-count rate, and zero reset time, this corresponds to an average waiting time of approximately $12$~ms per accepted state.
It is a heralding throughput over repeated discarded and accepted trials rather than the duration of a coherent two-qubit operation.
A successful state is created at the registered first click and corrected within the feedback latency.

Under those assumptions, the conditional fidelity is limited mainly by qubit dephasing, unmonitored cavity loss, finite click-time resolution, and feedback latency, whereas the throughput is limited by the kilohertz-scale spin-phonon coupling.
Increasing $T_2$ and reducing the click-to-feedback latency would primarily improve the fidelity; stronger confinement, a smaller acoustic mode volume, and improved qubit placement would primarily improve the rate.
The present result should be interpreted as probabilistic initialization of an odd-parity entangled state at the beginning of an experimental sequence.
Once heralded, that state can be used for tomography, a storage or coherence test, or a short locally controlled evolution.
The current coupling and rate do not support an on-demand entangling gate or a mid-circuit operation.

\section{Conclusion}
\label{Sec_6_conclusion}
We have established a direction-resolved microscopic theory of Rayleigh-type acoustic coupling to gate-defined Ge hole-spin qubits that treats coherent SAW driving and single-phonon interactions within the same acoustic modulation of the qubit $g$-matrix. The directional response arises because normal and sagittal-shear strain contributions transform differently under propagation reversal and interfere in the acoustic Larmor vector. This selectivity is electrically programmable, with the magnetic-field orientation setting a global Larmor reference.
Local dot-shape gates modify the static and acoustic $g$-matrix projections, continuously tuning the chirality and reversing the bright and nearly dark branches.
This local control leaves the acoustic circuit unchanged as the preferred propagation direction is reversed.

Quantizing the strain field yields a generalized anisotropic Rabi interaction whose co-rotating, counter-rotating, and longitudinal couplings are fixed by the acoustic field, magnetic-field orientation, and confinement.
For reciprocal traveling modes, the classical chirality measures the transverse quantum anisotropy, $\mathcal{A}_\pm=\pm\chi$, and equals the directional one-phonon emission contrast for equally normalized branches.

The bichromatic coherence protocol exploits this gate-tunable chirality. With the acoustic controls fixed after calibration, the selected working point has $\chi=0.996$, a $20$-fold contrast in Rabi frequency, and a $470$-fold contrast in transition strength.
The second-dressing resonance therefore addresses only the bright branch, while the two successive dressings provide the noise protection. The coherence time increases from $T_{2}^*=7.19~\mu\mathrm{s}$ for bare free induction ($T_{2,1}^*=5.40~\mu\mathrm{s}$ under a single dressing) to $T_{2,2}^*=167~\mu\mathrm{s}$ under bichromatic dressing.

The PnC design remains in the weak-coupling regime, with a local defect-mode coupling of order $1$~kHz.
We therefore use it as a monitored initialization channel rather than an on-demand gate.
The first registered phonon prepares a phase-defined odd-parity state with $F_{\text{Bell}}^{\text{cond}}\simeq0.88$ at an accepted rate $R_{\text{acc}}\simeq82~\mathrm{s}^{-1}$.
This corresponds to an average waiting time of about $12~\mathrm{ms}$ under ideal detection and zero reset time. The same framework extends naturally to multiqubit devices sharing acoustic control with local gate programmability.

Looking ahead, buried unstrained Ge channels offer a promising route to stronger acoustic control~\cite{Costa2026Buried,Mauro2025Unstrained}. Removing the large strain-induced contribution to the HH-LH splitting enhances band mixing and can increase the strain-induced $g$-matrix response. This suggests that both SAW-driven spin rotations and single-phonon coupling could be substantially enhanced. The larger deformation-potential-mediated spin-relaxation rates predicted for unstrained Ge support this prospect~\cite{Mauro2025Unstrained}, while also highlighting the need to control unwanted phonon channels. Joint engineering of the heterostructure and acoustic environment could therefore extend the present framework toward more efficient coherent spin-phonon interfaces.

\begin{acknowledgments}
Work supported by the Spanish Ministry of Science, Innovation, and Universities through Grants RYC2022-037527-I and PID2023-148257NA-I00 funded by MICIU/AEI/10.13039/501100011033 and by FSE+ and FEDER, UE, ``ERDF A way of making Europe'' and European Union Next Generation EU/PRTR.
This work is also supported by the Spanish National Plan for Quantum Communications under project COMQ2025001. JCAU, RSS, and DR acknowledge the support of the CSIC's Quantum Technologies Platform (QTEP) and the Severo Ochoa Centers of Excellence program through Grant CEX2024-001445-S. RSS acknowledges funding from the Spanish Comunidad de Madrid (CM) ``Talento Program'' (Project No. 2022-T1/IND-24070). IC acknowledges support from the Spanish Ministry of Universities through the FPU fellowship FPU20/03180. N.E.R. acknowledges the support of a fellowship from the “la Caixa” Foundation (ID 100010434). The fellowship code is LCF/BQ/DFR25/12000075.
\end{acknowledgments}

\appendix
\section{Multiband model, qubit projection, and strain selection rules}
\label{App:QubitProjection}
This appendix specifies the six-band Luttinger-Kohn-Bir-Pikus model and the projection used to obtain the qubit Hamiltonian in Eq.~\eqref{Eq:qubit_hamiltonian}.
The confined-hole Hamiltonian is:
\begin{equation}
H = H_{\text{LK}}^{(6)} + H_{\text{BP}} + H_Z + V_{\text{conf}}(\bm{x}) + V_E(\bm{x}) \,,
\end{equation}
where $\bm{x}=(x,y,z)$ and the growth direction is $z$.
The ordered Bloch basis is $\ket{3/2,3/2}$, $\ket{3/2,-3/2}$, $\ket{3/2,1/2}$, $\ket{3/2,-1/2}$, $\ket{1/2,1/2}$, $\ket{1/2,-1/2}$, containing the heavy-hole, light-hole, and split-off components.
In this basis, the six-band Luttinger-Kohn Hamiltonian is:
\begin{equation}
H_{\text{LK}}^{(6)} = \mqty[H_{\text{HL}} & W \\ W^\dagger & (P+\Delta_{\text{SO}})\mathbb{1}_2] \,,
\end{equation}
where the heavy-hole-light-hole block is:
\begin{equation}
H_{\text{HL}} = \mqty[
P+Q & 0 & -S & R \\
0 & P+Q & R^\dagger & S^\dagger \\
-S^\dagger & R & P-Q & 0 \\
R^\dagger & S & 0 & P-Q
] \,,
\end{equation}
and its coupling to the split-off sector is:
\begin{equation}
W = \mqty[
-\frac{S}{\sqrt{2}} & \sqrt{2}R \\
-\sqrt{2}R^\dagger & -\frac{S^\dagger}{\sqrt{2}} \\
-\sqrt{2}Q & \sqrt{\frac{3}{2}}S \\
\sqrt{\frac{3}{2}}S^\dagger & \sqrt{2}Q
] \,,
\end{equation}
where $\Delta_{\text{SO}}$ is the split-off energy.
The kinetic blocks are:
\begin{equation}
\begin{aligned}
P &= \frac{\hbar^2\gamma_1}{2m_0}(k_x^2+k_y^2+k_z^2) \,,\\
Q &= \frac{\hbar^2\gamma_2}{2m_0}(k_x^2+k_y^2-2k_z^2) \,,\\
R &= \frac{\sqrt{3}\hbar^2}{2m_0}[-\gamma_2(k_x^2-k_y^2)+2\mathrm{i}\gamma_3\{k_x,k_y\}] \,,\\
S &= \frac{\sqrt{3}\hbar^2\gamma_3}{m_0}[\{k_x,k_z\}-\mathrm{i}\{k_y,k_z\}] \,,
\end{aligned}
\end{equation}
with $\{k_\alpha,k_\beta\}=(k_\alpha k_\beta+k_\beta k_\alpha)/2$.
Orbital magnetic coupling is retained through $k_\alpha=-\mathrm{i}\pdv*{}{x_\alpha}+eA_\alpha(\bm{x})/\hbar$, with $\bm{B}=\nabla\times\bm{A}$.

The Bir-Pikus Hamiltonian has the same six-band block structure, with $\Delta_{\text{SO}}=0$ and the kinetic blocks replaced by:
\begin{equation}
\begin{aligned}
P_\epsilon &= -a_v(\epsilon_{xx}+\epsilon_{yy}+\epsilon_{zz}) \,,\\
Q_\epsilon &= -\frac{b_v}{2}(\epsilon_{xx}+\epsilon_{yy}-2\epsilon_{zz}) \,,\\
R_\epsilon &= \frac{\sqrt{3}b_v}{2}(\epsilon_{xx}-\epsilon_{yy})-\mathrm{i}d_v\epsilon_{xy} \,,\\
S_\epsilon &= -d_v(\epsilon_{xz}-\mathrm{i}\epsilon_{yz}) \,,
\end{aligned}
\end{equation}
where $a_v$, $b_v$, and $d_v$ are the deformation potentials. The full six-band Zeeman matrix is parameterized by $\kappa$ and $q$, denoted $\kappa_1$ and $\kappa_2$, respectively, in the numerical implementation, and is retained together with the orbital magnetic coupling~\cite{Luttinger1955,Luttinger1956,Winkler2003}.

The vertical envelope is expanded in a symmetric infinite-well basis of width $L_z$, while the lateral confinement is harmonic:
\begin{equation}
V_{\text{conf}}(x',y') = \frac{1}{2}m_\parallel\omega_{x'}^2x'^2 + \frac{1}{2}m_\parallel\omega_{y'}^2y'^2 \,,
\end{equation}
and the vertical electric field enters through $V_E=eE_z z$.
The axes $x'$ and $y'$ are rotated by $\phi_{\text{dot}}$ relative to the crystal axes, and the confinement lengths are parameterized by $a_{x'}=a_0/\sqrt{\alpha}$ and $a_{y'}=a_0\sqrt{\alpha}$.
The numerical basis retains four states along each spatial direction, giving a Hamiltonian dimension of $6\times4^3=384$.
The parameters used unless scanned are listed in Table~\ref{tab:parameters}.

After diagonalization, the lowest Zeeman-split doublet defines $\ket{0}$ and $\ket{1}$ and the projector $P_Q=\ketbra{0}{0}+\ketbra{1}{1}$.
After subtracting the scalar energy $E_Q$, the static projection is:
\begin{equation}
P_Q H_0 P_Q-E_Q\mathbb{1} = \frac{\mu_B}{2}\bm{\sigma}\cdot\tilde{g}_0\bm{B} = \frac{\hbar}{2}\bm{\Omega}_0\cdot\bm{\sigma} \,,
\end{equation}
and the positive-frequency acoustic vertex is projected as:
\begin{equation}
\label{Eq:projected_vertex}
\begin{aligned}
V_{\xi,+} &= \sum_{\mu\nu}\epsilon^+_{\mu\nu,\xi}D_{\mu\nu} \,, \\
P_Q V_{\xi,+}P_Q-V_{I,\xi}\mathbb{1} &= \frac{\mu_B}{2}\bm{\sigma}\cdot\delta\tilde{g}_{\xi,+}\bm{B} = \frac{\hbar}{2}\delta\bm{\Omega}_{\xi,+}\cdot\bm{\sigma} \,,
\end{aligned}
\end{equation}
where $D_{\mu\nu}=\pdv*{H_{\text{BP}}}{\epsilon_{\mu\nu}}$ and $V_{I,\xi}$ is a scalar shift.

For a heavy-hole-dominated qubit, $R_\epsilon$ couples heavy-hole states to the opposite-sign light-hole components and produces the leading in-plane strain response, whereas $S_\epsilon$ couples them to the same-sign light-hole components and produces the sagittal-shear response.
The diagonal blocks $P_\epsilon$ and $Q_\epsilon$ primarily shift the heavy-hole-light-hole splitting.
Equation~\eqref{Eq:g_corrections} summarizes these leading channels, while all numerical results use the full six-band eigenstates and finite-momentum Bir-Pikus matrix elements.
\begin{table}[H]
\caption{Baseline parameters used for the six-band qubit calculations, unless the corresponding quantity is scanned.}
\label{tab:parameters}
\begin{ruledtabular}
\begin{tabular}{ccc}
Parameter & Value & Note \\
\hline
$a_{x'},a_{y'},L_z$ & $40,40,11$~nm & Quantum dot size \\
$n_z,n_x,n_y$ & $4,4,4$ & Basis truncation \\
$E_z$ & $1$~MV/m & Gate electric field \\
$\theta_B$ & $45^\circ$ & Planar $\bm{B}$ angle \\
$f_{\text{SAW}}$ & $2$~GHz & Qubit frequency
\end{tabular}
\end{ruledtabular}
\end{table}
We checked the sensitivity to the band content by repeating the calculations underlying Figs.~\ref{Fig_2}-\ref{Fig_6} with an eight-band model that additionally includes the $\Gamma_6$ conduction doublet.
Across the full angular maps, the eight-band calculation preserves the dominant structure, with maximum deviations from the six-band results below $0.004$ in $\Lambda_{\text{TL}}$ for Fig.~\ref{Fig_2} and $0.013$ in $\chi$ for Fig.~\ref{Fig_3}.
At the working points used in the applications, the chirality and coherence enhancement differ by less than $1\%$, while the Bell-state fidelity differs by less than $10^{-5}$.
The six-band model is therefore sufficient for the conclusions reported here.

\section{Rayleigh mode conventions, propagation parity, and classical drive}
\label{App:RayleighMode}
This appendix fixes the analytical Rayleigh conventions used in Sec.~\ref{Sec_2_model} and connects the finite-momentum Bir-Pikus projection to the $g$-matrix rates in Eq.~\eqref{Eq:gmatrix_fTL}.
We denote the coordinate along the propagation axis by $x_\parallel=\bm{x}\cdot\hat{\bm{e}}$.
The positive-frequency Rayleigh displacement in this coordinate is written in the form:
\begin{equation}
\bm{u}^+_\xi(\bm{x}) = \frac{U_0}{2}[-\xi A_\parallel(z)\mathrm{e}^{\mathrm{i}\delta}\hat{\bm{e}}+A_z(z)\hat{\mathbf{z}}]\mathrm{e}^{\mathrm{i}\xi kx_\parallel} \,.
\end{equation}
Here $U_0$ is the surface displacement amplitude, and $A_\parallel(z)$ and $A_z(z)$ are real depth profiles.
The parameter $\delta$ is the relative phase between the in-plane and out-of-plane Rayleigh motion.
The superscript $+$ denotes the coefficient of $\mathrm{e}^{-\mathrm{i}\omega t}$.
For a physical eigenmode, the profiles and their relative phase are fixed by the elastic boundary conditions.

With the strain tensor $\epsilon_{\mu\nu}=(\pdv*{u_\nu}{x_\mu}+\pdv*{u_\mu}{x_\nu})/2$, the displacement above gives the following depth profiles:
\begin{equation}
\begin{aligned}
\eta_1(z) &= -\mathrm{i}kA_\parallel(z)\mathrm{e}^{\mathrm{i}\delta} \,,\\
\eta_2(z) &= \frac{1}{2}\left[-\dv{A_\parallel(z)}{z}\mathrm{e}^{\mathrm{i}\delta}+\mathrm{i}kA_z(z)\right] \,,\\
\eta_3(z) &= \dv{A_z(z)}{z} \,.
\end{aligned}
\end{equation}
The corresponding positive-frequency strain profiles in the SAW frame are then:
\begin{equation}
\label{Eq:saw_strain_positive}
\bm{\epsilon}^+_\xi(\bm{x}) = \frac{U_0}{2}\mathrm{e}^{\mathrm{i}\xi kx_\parallel}(\eta_1(z),\xi\eta_2(z),\eta_3(z)) \,.
\end{equation}
Here $\eta_1$, $\eta_2$, and $\eta_3$ are the direction-independent profiles associated with $\epsilon_{\parallel\parallel}$, $\epsilon_{\parallel z}$, and $\epsilon_{zz}$, respectively.
Equation~\eqref{Eq:saw_strain_positive} makes the propagation parity explicit: the normal-strain profiles are even in $\xi$, whereas the sagittal shear is odd.

After suppressing the common factor $(U_0/2)\mathrm{e}^{\mathrm{i}\xi kx_\parallel}$, rotation from the SAW frame to the crystal axes gives:
\begin{equation}
\label{Eq:saw-strain-rotation}
\begin{aligned}
\epsilon_{xx} &= \cos^2\theta_{\text{SAW}}\eta_1 \,, \\
\epsilon_{yy} &= \sin^2\theta_{\text{SAW}}\eta_1 \,, \\
\epsilon_{xz} &= \xi\cos\theta_{\text{SAW}}\eta_2 \,, \\
\epsilon_{yz} &= \xi\sin\theta_{\text{SAW}}\eta_2 \,, \\
\epsilon_{xy} &= \sin\theta_{\text{SAW}}\cos\theta_{\text{SAW}}\eta_1 \,, \\
\epsilon_{zz} &= \eta_3 \,.
\end{aligned}
\end{equation}
Equation~\eqref{Eq:saw-strain-rotation} reproduces Eq.~\eqref{Eq:rayleigh_to_crystal} when inserted into the strain combinations of Eq.~\eqref{Eq:g_corrections}.

The positive-frequency Bir-Pikus operator generated by Eq.~\eqref{Eq:saw_strain_positive} is:
\begin{equation}
V^{\text{C}}_{\xi,+} = \frac{U_0}{2}\mathrm{e}^{\mathrm{i}\xi kx_\parallel}\mathcal{M}_\xi \,,
\end{equation}
where the decomposition is:
\begin{equation}
\label{Eq:app_Mxi}
\mathcal{M}_\xi = \mathcal{M}_{\text{e}} + \xi\mathcal{M}_{\text{o}},\quad \mathcal{M}_{\text{e}} = D_1\eta_1 + D_3\eta_3,\quad \mathcal{M}_{\text{o}} = D_2\eta_2 \,,
\end{equation}
and its components are defined as:
\begin{equation}
\begin{aligned}
D_1 &= \cos^2\theta_{\text{SAW}}D_{xx} + \sin^2\theta_{\text{SAW}}D_{yy} \\
&\quad + \sin\theta_{\text{SAW}}\cos\theta_{\text{SAW}}D_{xy} \,, \\
D_2 &= \cos\theta_{\text{SAW}}D_{xz} + \sin\theta_{\text{SAW}}D_{yz} \,,\\
D_3 &= D_{zz} \,,
\end{aligned}
\end{equation}
where $D_{\mu\nu}=\pdv*{H_{\text{BP}}}{\epsilon_{\mu\nu}}$ as in Eq.~\eqref{Eq:projected_vertex}.
Thus $\mathcal{M}_{\text{e}}$ contains the normal-strain vertex and $\mathcal{M}_{\text{o}}$ contains the sagittal-shear vertex.
The preceding equations provide the ideal Rayleigh reduction used to identify the propagation-even and propagation-odd strain channels.

The projected transverse and longitudinal amplitudes are:
\begin{equation}
\begin{aligned}
v_\xi^\perp &= \mel*{1}{V^{\text{C}}_{\xi,+}}{0} \,,\\
v_\xi^\parallel &= \frac{\mel*{1}{V^{\text{C}}_{\xi,+}}{1}-\mel*{0}{V^{\text{C}}_{\xi,+}}{0}}{2} \,,
\end{aligned}
\end{equation}
where the labels are defined relative to the Larmor axis and $v_\xi^\parallel=(\hbar/2)\hat{\bm{n}}\cdot\delta\bm{\Omega}_{\xi,+}$.
For the ideal Rayleigh mode, Eq.~\eqref{Eq:app_Mxi} gives:
\begin{equation}
v_\xi^\perp = \frac{U_0}{2}\mel*{1}{\mathrm{e}^{\mathrm{i}\xi kx_\parallel}(\mathcal{M}_{\text{e}}+\xi\mathcal{M}_{\text{o}})}{0} = v_{\text{e}}^\perp+\xi v_{\text{o}}^\perp \,,
\end{equation}
with $v_{\text{e}}^\mu=(v_+^\mu+v_-^\mu)/2$ and $v_{\text{o}}^\mu=(v_+^\mu-v_-^\mu)/2$.

In the local limit $ka_{\text{dot}}\ll1$, the even and odd amplitudes reduce to the projected normal-strain and sagittal-shear amplitudes, respectively, whereas the finite plane-wave phase mixes the two sectors.
The chirality of Eq.~\eqref{Eq:chirality} becomes:
\begin{equation}
\chi = \frac{2\Re[(v_{\text{e}}^\perp)^*v_{\text{o}}^\perp]}{|v_{\text{e}}^\perp|^2+|v_{\text{o}}^\perp|^2} \,,
\end{equation}
which requires both propagation-even and propagation-odd amplitudes with a finite real overlap.

For the quantitative results at $\theta_{\text{SAW}}=0^\circ$, the classical drive is constructed from the complete FEM field.
The drive for other propagation axes is obtained by rotating this reference field:
\begin{equation}
V^{\text{C}}_{\xi,+} = \sum_{\mu\nu}D_{\mu\nu}\epsilon_{\mu\nu,\xi}^{\text{FEM},+}(\bm{x}) \,.
\end{equation}
The sum in this expression runs over the six independent symmetric strain components.
The complex FEM fields already contain the full spatial phase and do not require a separate plane-wave factor.

Using the quadrature convention of Eq.~\eqref{Eq:gmatrix_quadratures}, the rotating-wave rates associated with the projected amplitudes are:
\begin{equation}
f_{T,\xi} = \frac{2|v_\xi^\perp|}{h},\quad f_{L,\xi} = \frac{4|v_\xi^\parallel|}{h} \,.
\end{equation}
These expressions are equivalent to Eq.~\eqref{Eq:gmatrix_fTL}, with $v_\xi^\perp$ and $v_\xi^\parallel$ evaluated from the complete FEM Bir-Pikus operator in the numerical calculation.

\section{Finite element simulations}
\label{App:FiniteElementSimulations}
To quantify the SAW fields experienced by the quantum dots, we developed a 3D frequency-domain finite element model of the launcher–heterostructure region.
The model incorporates a $128^\circ$ YX LiNbO${}_3$/SiO${}_2$ piezoelectric launcher, the Ge/SiGe/Si heterostructure, the multilayer gate stack, and a quantum dot aligned with the SAW propagation direction as shown in Fig.~\ref{Fig_1}(a).
The open-launcher wavelength $\lambda_{\text{open}}$ and IDT periodicity at $2$~GHz were first calibrated from a piezoelectric semi-analytical finite element (SAFE) eigenmode calculation of the loaded Rayleigh-like mode.

We then solved the coupled elastodynamic and quasi-electrostatic equations, including the anisotropic material response, cryogenic-cooldown stress and the epitaxial pre-strain of the Ge quantum well, while perfectly matched layers absorbed outgoing waves.
Each quantum dot was treated as a sampling volume within the continuous Ge quantum well, rather than as a separate mechanical inclusion.
This allowed the complex displacement and all six strain components to be averaged locally, resolving their amplitudes and phases across the array.
\begin{figure*}[htbp!]
\centering
\includegraphics[width=\linewidth]{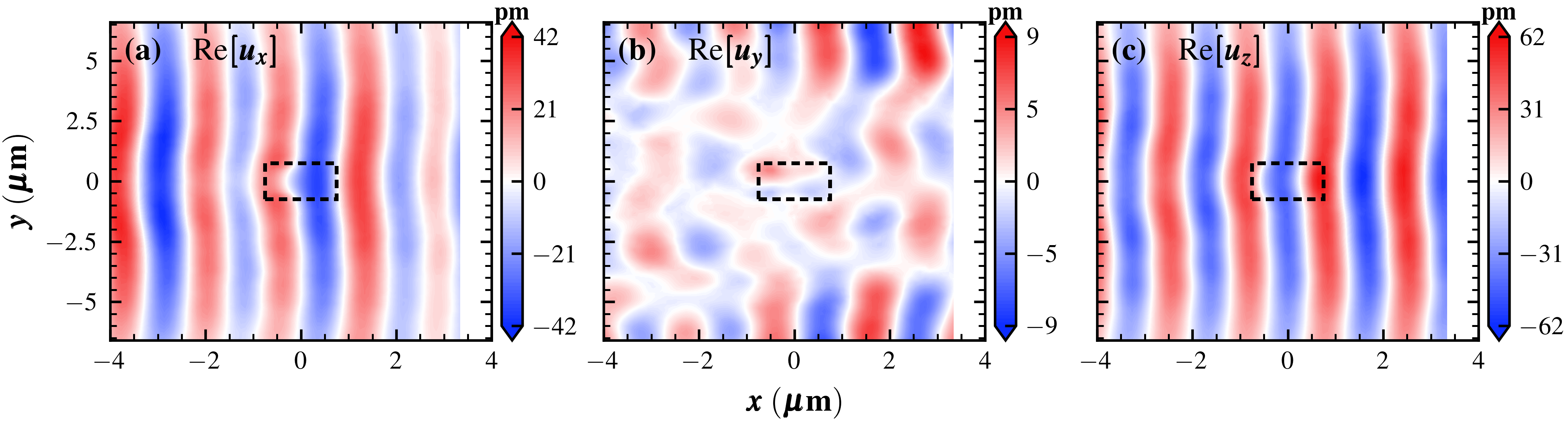}
\caption{Spatial distributions of the three displacement components for the single-qubit design, evaluated at $z=0$ (the quantum-dot plane).
The black dashed rectangles indicate the local device region centered at the origin. This figure provides a representative demonstration of the displacement field for the single-qubit design with perfectly matched layer boundary conditions.}
\label{Fig:AppendixFEMDisplacement}
\end{figure*}

The analytical Rayleigh wave solution provides a reference for the dominant polarization and phase relation between the longitudinal and vertical components.
In the full device, however, the SAW profile is reshaped by the heterogeneous and mechanically loaded environment.
These effects, most importantly, modify the relative amplitudes and phases of the displacement and strain components and generate additional shear strains. Representative displacement fields are shown in Fig.~\ref{Fig:AppendixFEMDisplacement}(a)-(c).

The FEM field contains a finite transverse displacement $u_y$, which is absent for an ideal isotropic Rayleigh wave. However, under the mirror operation $y\to-y$, a mirror-symmetric displacement field requires $u_y$ to vanish only on the mirror plane $y=0$.
The finite IDT aperture and local mechanical loading introduce transverse field gradients that allow a mirror-odd $u_y$ component.

In addition, the anisotropic elastic and piezoelectric response of $128^\circ$ YX LiNbO$_3$ weakly hybridizes the sagittal Rayleigh polarization with shear-horizontal motion, yielding a quasi-Rayleigh mode.
Consistent with this interpretation, $|u_y|$ reaches only approximately $9$~pm per applied volt at the IDT, compared with approximately $42$~pm and $62$~pm per applied volt for $|u_x|$ and $|u_z|$, respectively.
The finite $u_y$ represents symmetry-allowed transverse motion and weak anisotropy-induced mode mixing, while the overall polarization remains predominantly Rayleigh-like.

Away from the dashed device region, $u_x$ and $u_z$ exhibit nearly regular oscillations along $x$ with only weak variation along $y$.
This indicates that the displacement field remains approximately uniform across the transverse direction.
Within and near the device region, however, the oscillatory pattern is locally distorted and the displacement amplitude is reduced.
This suppression is consistent with the greater effective stiffness and mechanical loading of the device stack, which constrain the local motion and redistribute the displacement field without eliminating the overall oscillatory pattern.

\section{Noise model and fixed-input two-IDT Floquet coherence}
\label{App:BichromaticCoherence}
This appendix specifies the noise spectra, relaxation rates, and numerical procedure used to obtain the fixed-input two-IDT coherence results in Fig.~\ref{Fig_5}.
The rotating-frame Hamiltonians and nested-dressing hierarchy are given in Eqs.~\eqref{Eq:bichromatic_first_frame}-\eqref{Eq:bichromatic_noise_hierarchy} and are not repeated here.
All frequencies below are ordinary frequencies rather than angular frequencies.
Unless stated otherwise, the power spectral densities (PSDs) are one-sided, the offset frequency satisfies $f>0$, and frequency-noise PSDs are expressed in units of Hz$^2$/Hz.

For frames $m=0,1,2$, corresponding to the bare, first-dressed, and second-dressed qubits, respectively, let $\delta\nu_m(t)$ denote the effective longitudinal fluctuation of the instantaneous splitting about its mean.
The accumulated random phase is:
\begin{equation}
\phi_m(t) = 2\pi\int_0^t \delta\nu_m(t') \dd{t'} \,.
\end{equation}
The dephasing factor is $\mathrm{e}^{-K_m(t)}=\expval**{\mathrm{e}^{-\mathrm{i}\phi_m(t)}}$.
Including population relaxation, the normalized coherence envelope is:
\begin{equation}
C_m(t) = \mathrm{e}^{-K_m(t)}\mathrm{e}^{-\Gamma_{1,m}^{\text{eff}}t/2} \,.
\end{equation}
Within the stationary Gaussian filter-function treatment used here, the one-sided longitudinal frequency-noise spectrum $S_{\nu,m}(f)$ gives:
\begin{equation}
K_m(t) = \frac{(2\pi)^2}{2}\int_{f_{\text{lo}}}^{f_{\text{hi}}}S_{\nu,m}(f)\left[\frac{\sin(\pi f t)}{\pi f}\right]^2 \dd{f} \,.
\end{equation}
Thus, $K_m(t)$ is calculated from the frame-dependent spectrum $S_{\nu,m}(f)$ and is not an independent fitting function.
The coherence time $T_{2,m}^{\ast}$ is defined by $C_m(T_{2,m}^{\ast})=\mathrm{e}^{-1}$.

The calculation uses a circular dot with $\alpha=1$ and an acoustic propagation angle $\theta_{\text{SAW}}=4^\circ$.
The primary IDT launches the $f_1=2$~GHz mode along $\xi_1=+1$, whereas the counterpropagating auxiliary IDT launches the $f_2=50$~MHz mode along $\xi_2=-1$.
At every magnetic field orientation $\theta_B$, the field magnitude is adjusted so that $f_q(\theta_B)=f_1$.
The FEM-LKBP calculation supplies the voltage-normalized primary transverse coupling $G_{T1,+}(\theta_B)$, the power-normalized auxiliary longitudinal coupling $G_{L2,-}(\theta_B)$, and the chirality $\chi(\theta_B)$. The auxiliary longitudinal coupling $G_{L2,-}$ is obtained from an independent $50$~MHz layered piezoelectric SAFE-mode calculation.

The primary co-propagating rate is $f_{R1,+}(\theta_B)=V_1G_{T1,+}(\theta_B)$.
The counterpropagating rate is $f_{R1,-}(\theta_B)=f_{R1,+}(\theta_B)\sqrt{[1-\chi(\theta_B)]/[1+\chi(\theta_B)]}$.
The auxiliary longitudinal rate is $f_{R2}(\theta_B)=G_{L2,-}(\theta_B)\sqrt{P_2}/2$.
The second-frame detuning is $\Delta_2(\theta_B)=f_{R1,+}(\theta_B)-f_2$.
Here $V_1$ is the differential peak voltage applied to the primary IDT and $P_2$ is the acoustic line power of the auxiliary mode.
The factor $1/2$ in $f_{R2}=f_{L2,-}/2$ follows from the rotating-wave reduction in Eq.~\eqref{Eq:bichromatic_second_frame}.

The electric-field susceptibilities are evaluated along the resonant $f_q=2$~GHz trajectory.
At each magnetic field orientation, the qubit frequency is calculated at 11 equally spaced values over $E_z^{(0)}-0.05$~MV/m $\leq E_z\leq E_z^{(0)}+0.05$~MV/m.
A fourth-order polynomial fit to $f_q(E_z)-f_q(E_z^{(0)})$ defines $\beta_1=\eval{\pdv*{f_q}{E_z}}_{E_z^{(0)}}$ and $\beta_2=\eval{\pdv*[2]{f_q}{E_z}}_{E_z^{(0)}}$.
The fit is repeated over an inner $\pm0.03$~MV/m window to monitor the convergence of $\beta_2$.

With $f_0=1$~Hz, the vertical electric-field noise is modeled as $S_{E_z}(f)=A_{E_z}^2(f/f_0)^{-\alpha_E}$.
The resulting qubit-frequency fluctuation is:
\begin{equation}
\delta f_q(t) = \beta_1\delta E_z(t)+\frac{\beta_2}{2}\left[\delta E_z^2(t)-\expval**{\delta E_z^2}\right]+\delta f_{\text{HF}}(t) \,.
\end{equation}
The linear charge-noise spectrum is $S_{E,\text{lin}}(f)=\beta_1^2S_{E_z}(f)$.
For Gaussian electric-field noise, the slow variance, rms quadratic shift, and effective quadratic spectrum are:
\begin{align}
\sigma_{E_z}^2 &= \int_{f_{\text{lo}}}^{f_{\text{slow}}}S_{E_z}(f) \dd{f} \,,\\
\sigma_{E,\text{quad}} &= \frac{\abs{\beta_2}\sigma_{E_z}^2}{\sqrt{2}} \,,\\
S_{E,\text{quad}}(f) &= S_{1/f}(f;\sigma_{E,\text{quad}}) \,.
\end{align}
The amplitude $A_{E_z}$ is calibrated at $\theta_B=45^\circ$ by setting $K_0(10~\mu\mathrm{s})=1$ with only $S_{E,\text{lin}}+S_{E,\text{quad}}$ included, before adding hyperfine noise and relaxation.

The natural-Ge hyperfine contribution is $S_{\text{HF}}(f)=S_{0,\text{HF}}(f/f_0)^{-\alpha_{\text{HF}}}$.
The bare longitudinal spectrum is $S_{\nu,0}(f)=S_{E,\text{lin}}(f)+S_{E,\text{quad}}(f)+S_{\text{HF}}(f)$.
For each acoustic tone $j=1,2$, the Rabi-frequency fluctuation is $\delta f_{Rj}(t)=f_{Rj}a_j(t)$, where $a_j(t)$ is the fractional amplitude fluctuation.
Its band-limited spectrum is:
\begin{equation}
S_{a_j}(f) = \eta_{Aj}^2\frac{\Theta(f-f_{A,\text{lo}})\Theta(f_{A,\text{hi}}-f)}{f\ln(f_{A,\text{hi}}/f_{A,\text{lo}})} \,.
\end{equation}
This normalization gives an integrated variance $\eta_{Aj}^2$ over the amplitude-noise bandwidth.

The source single-sideband phase noise is modeled as $L_\phi(f)=L_{\phi,0}(f/f_{\phi,0})^{s_\phi}$. In the numerical calculation, $L_{\phi,0}$ is converted from dBc/Hz to linear units before evaluating $S_\phi(f)=2L_\phi(f)$.
With the one-sided convention $S_\phi(f)=2L_\phi(f)$, the instantaneous source-frequency-noise spectrum is $S_d(f)=f^2S_\phi(f)$, consistent with $2\pi\delta f_d(t)=\dv*{\phi(t)}{t}$.
The phenomenological acoustic-path jitter is:
\begin{equation}
S_{\text{jit}}(f) = \frac{\sigma_{f,\text{jit}}^2\Theta(f-f_{J,\text{lo}})\Theta(f_{J,\text{hi}}-f)}{f\ln(f_{J,\text{hi}}/f_{J,\text{lo}})} \,.
\end{equation}
The primary source noise and its path jitter enter the first-frame detuning through $S_d(f)+S_{\text{jit}}(f)$.
The residual source-frequency and differential path noise entering the second-frame detuning is controlled by $\epsilon_{\text{rel}}=\max[\epsilon_0,2(1-\rho_{12})]$.
Its spectrum is $S_{2,\text{rel}}(f)=\epsilon_{\text{rel}}S_d(f)+2S_{\text{jit}}(f)$.
The common-source contribution is reduced by the correlation coefficient $\rho_{12}$, while the two path-jitter contributions are treated as independent.

The same quasistatic construction is used for transverse noise in either dressed frame.
For a transverse spectrum $S_\perp(f)$ and an ordinary-frequency dressed splitting $F$, we define:
\begin{align}
\sigma_\perp^2(F) &= \int_{f_{\text{lo}}}^{f_c(F)}S_\perp(f) \dd{f} \,,\\
f_c(F) &= \min\left[f_{\text{slow}},\max\left(2f_{\text{lo}},\frac{F}{5}\right)\right] \,,\\
\sigma^{(2)}(F) &= \frac{\sigma_\perp^2(F)}{\sqrt{2}F} \,,\\
S_{\perp,\text{eff}}^{(2)}(f;F) &= S_{1/f}(f;\sigma^{(2)}(F)) \,.
\end{align}
This effective spectrum has the same integrated variance as the quadratic frequency shift generated by the slow transverse fluctuations.

The noise transverse to the first-dressed quantization axis is $S_{\Delta_1}(f)=S_{\nu,0}(f)+S_d(f)+S_{\text{jit}}(f)$.
Applying the preceding construction with $S_\perp(f)=S_{\Delta_1}(f)$ and $F=f_{R1,+}$ gives $S_{\Delta_1,\text{eff}}^{(2)}(f)$.
The first-dressed longitudinal spectrum is $S_{\nu,1}(f)=f_{R1,+}^2S_{a_1}(f)+S_{\Delta_1,\text{eff}}^{(2)}(f)$.
The associated relaxation rate is $\Gamma_{1\rho}=(2\pi)^2S_{\Delta_1}(f_{R1,+})/2$.

Away from the matching angle, the auxiliary mode drives a detuned second-dressed transition.
The effective second-dressed splitting is $F_2=\sqrt{\Delta_2^2+f_{R2}^2}$, with direction cosines $c_2=\Delta_2/F_2$ and $s_2=f_{R2}/F_2$.
The first-dressed detuning noise entering this tilted basis is $S_{\Delta_2}(f)=S_{\nu,1}(f)+S_{2,\text{rel}}(f)$.
The corresponding auxiliary amplitude noise is $S_{A2}(f)=f_{R2}^2S_{a_2}(f)$.
Assuming these contributions are uncorrelated, their first-order longitudinal and transverse projections are:
\begin{align}
S_{\parallel,2}^{(1)}(f) &= c_2^2S_{\Delta_2}(f)+s_2^2S_{A2}(f) \,,\\
S_{\perp,2}(f) &= s_2^2S_{\Delta_2}(f)+c_2^2S_{A2}(f) \,.
\end{align}
Applying the quasistatic construction with $S_\perp(f)=S_{\perp,2}(f)$ and $F=F_2$ gives $S_{\perp,2,\text{eff}}^{(2)}(f)$.
The two-IDT longitudinal spectrum is $S_{\nu,2}(f)=S_{\parallel,2}^{(1)}(f)+S_{\perp,2,\text{eff}}^{(2)}(f)$.
The corresponding relaxation rate is $\Gamma_{1\rho\rho}=(2\pi)^2S_{\perp,2}(F_2)/2$.
At $\Delta_2=0$, the tilted-axis expressions reduce to the resonant second-dressed model used during the single-point calibration.

The laboratory-frame relaxation rate is $\Gamma_{\text{lab}}=\Gamma_{\text{bg}}+(2n_{\text{th}}+1)\Gamma_{\text{SAW}}^{(0)}$, where $\Gamma_{\text{bg}}=1/T_1^{\text{bg}}$ and $n_{\text{th}}=[\mathrm{e}^{hf_1/(k_{\text{B}}T_{\text{eff}})}-1]^{-1}$.
The zero-temperature rate into the two counterpropagating Rayleigh-SAW continua is estimated from the FEM transverse coupling and the local vertical displacement as:
\begin{equation}
\label{Eq:AppendixSAWRelaxation}
\Gamma_{\text{SAW}}^{(0)} = \sum_{\xi=\pm1}\frac{\hbar}{2\rho_{\text{Ge}}L_y\omega_1I_{k_1}v_{\text{SAW}}}\abs{\frac{2\pi f_{R1,\xi}}{u_z}}^2 \,.
\end{equation}
Here $\omega_1=2\pi f_1$, $k_1=\omega_1/v_{\text{SAW}}$, and the Rayleigh depth-normalization integral is:
\begin{small}
\begin{equation}
I_{k_1} = \frac{\left[\frac{2(1+q^2)}{q}-\frac{4(1+q^2)}{p}+\frac{(1+q^2)^2(1+p^2)}{2p^3}\right]}{k_1(1-q^2)^2} \,.
\end{equation}
\end{small}
The dimensionless parameters $p$ and $q$ describe the two Rayleigh depth-decay scales.

The quantities $f_{R1,\xi}$ and $u_z$ in Eq.~\eqref{Eq:AppendixSAWRelaxation} use the same classical-drive normalization, so their ratio is independent of $V_1$.
The value $\abs{u_z}/V_{\text{peak}}=43.7083$~pm/V is the modulus of the complex vertical-displacement phasor averaged over the quantum-dot volume.
It is distinct from the pointwise maximum of its real part at $z=0$ displayed in Appendix~\ref{App:FiniteElementSimulations}.
The effective bare-frame rate entering Eq.~\eqref{Eq:coherence} is $\Gamma_{1,0}^{\text{eff}}=\Gamma_{\text{lab}}$.
The corresponding first-dressed rate is $\Gamma_{1,1}^{\text{eff}}=\Gamma_{\text{lab}}+\Gamma_{1\rho}$.
The second-dressed rate is $\Gamma_{1,2}^{\text{eff}}=\Gamma_{\text{lab}}+\Gamma_{1\rho}+\Gamma_{1\rho\rho}$.

The two IDT inputs are calibrated only at the target angle $\theta_B^*=20^\circ$.
The primary voltage is fixed by $V_1=f_2/G_{T1,+}(\theta_B^*)$.
This gives $V_1=0.1031~\mathrm{V}_{\text{peak}}$ and $f_{R1,+}(\theta_B^*)=f_2=50$~MHz.
At the same angle, $f_{R2}$ is sampled at 280 logarithmically spaced points from $0.05$~MHz to $8$~MHz.
The slow drive-detuning variances used only to screen this calibration scan are:
\begin{align}
\sigma_{d,1}^2 &= \int_{f_{\text{lo}}}^{f_{\text{slow}}}[S_d(f)+S_{\text{jit}}(f)] \dd{f} \,,\\
\sigma_{d,2}^2 &= \int_{f_{\text{lo}}}^{f_{\text{slow}}}S_{2,\text{rel}}(f) \dd{f} \,.
\end{align}
With $\Gamma_{2,0}=1/T_{2,0}^*$ and $\Gamma_{2,1}=1/T_{2,1}^*$, the accepted gaps satisfy:
\begin{align}
f_{R1,+} &> \kappa_v\max\{\Gamma_{2,0},\Gamma_{\text{lab}},\sigma_{d,1},f_{\text{op}}^{\text{min}}\} \,,\\
f_{R2} &> \kappa_v\max\{\Gamma_{2,1},\Gamma_{\text{lab}}+\Gamma_{1\rho},\sigma_{d,2},f_{\text{op}}^{\text{min}}\} \,.
\end{align}
The primary gap is also restricted by $f_{R1,+}<0.25f_1$.
Maximizing $T_{2,2}^*$ over the accepted calibration points gives $f_{R2}(\theta_B^*)=1.274$~MHz.
The fixed auxiliary line power is $P_2=[2f_{R2}(\theta_B^*)/G_{L2,-}(\theta_B^*)]^2$, which gives $P_2=1.596$~mW/m.

After this single calibration, $V_1$, $P_2$, $f_1$, $f_2$, $\xi_1$, $\xi_2$, and $\theta_{\text{SAW}}$ remain fixed.
The final calculation evaluates $\theta_B$ and updates only the angle-dependent FEM-LKBP couplings, electric-field susceptibilities, $f_{R2}$, $\Delta_2$, $F_2$, and associated noise projections.
At fixed $P_2$, the value of $f_{R2}(\theta_B)$ changes only through $G_{L2,-}(\theta_B)$ and is not independently optimized.
No propagation direction, voltage, power, or dressed gap is reoptimized along this scan.
The smooth two-IDT curve in Fig.~\ref{Fig_5}(b) uses shape-preserving interpolation for display only and does not alter the 21 calculated points.
The fixed-control, noise, relaxation, and numerical parameters are summarized in Table~\ref{Tab:BichromaticNoiseParameters}.
\begin{table}[htbp!]
\caption{Parameters used for the fixed-input two-IDT coherence calculation in Fig.~\ref{Fig_5}.
The charge-noise amplitude is calibrated by imposing a charge-only coherence time of $10~\mu\mathrm{s}$ at $\theta_B=45^\circ$.}
\label{Tab:BichromaticNoiseParameters}
\begin{ruledtabular}
\begin{tabular}{cc}
Parameter & Value \\\hline
$\alpha,\theta_{\text{SAW}}$ & $1,4^\circ$ \\
$\theta_B^*,(\xi_1,\xi_2)$ & $20^\circ,(+1,-1)$ \\
$f_1,f_2$ & $2$~GHz, $50$~MHz \\
$V_1,P_2$ & $0.103064~\mathrm{V}_{\text{peak}}$, $1.59606$~mW/m \\
$f_{R2}(\theta_B^*)$ & $1.27404$~MHz \\
$A_{E_z}(f_0)$ & $0.154686$~(MV/m)/$\sqrt{\text{Hz}}$ \\
$\alpha_E,\alpha_{\text{HF}}$ & $1,1$ \\
$S_{0,\text{HF}}$ & $2.5\times10^6$~Hz$^2$/Hz \\
$\eta_{A1},\eta_{A2}$ & $10^{-3},10^{-3}$ \\
$f_{A,\text{lo}},f_{A,\text{hi}}$ & $1$~Hz, $10$~MHz \\
$10\log_{10}[L_{\phi,0}],f_{\phi,0},s_\phi$ & $-132.5$~dBc/Hz, $10$~kHz, $-2$ \\
$\sigma_{f,\text{jit}}$ & $50$~Hz \\
$f_{J,\text{lo}},f_{J,\text{hi}}$ & $1$~Hz, $10$~kHz \\
$\rho_{12},\epsilon_0$ & $0.99,0.01$ \\
$T_{\text{eff}},T_1^{\text{bg}}$ & $50$~mK, $10$~ms \\
$\rho_{\text{Ge}},L_y$ & $5323$~kg/m$^3$, $1$~$\mu$m \\
$p,q,v_{\text{SAW}}$ & $0.8475,0.3933,3978$~m/s \\
$\abs{u_z}/V_{\text{peak}}$ & $43.7083$~pm/V \\
$f_{\text{lo}},f_{\text{hi}}$ & $1$~Hz, $5$~GHz \\
$f_{\text{slow}},N_f$ & $1$~MHz, $2200$ \\
$\kappa_v,f_{\text{op}}^{\text{min}}$ & $5,10$~kHz \\
\end{tabular}
\end{ruledtabular}
\end{table}

\section{Fixed phononic-crystal cavity and click-resolved Bell-state initialization dynamics}
\label{App:BellStateDynamics}
This appendix specifies the fixed PnC cavity, the conversion from the direction-resolved open-SAW couplings to the local cavity couplings, and the counted dynamics used for the timestamp-corrected Bell-state initialization results of Fig.~\ref{Fig_6}.
The final calculation combines one fixed PnC cavity design with the angle-dependent complex open-SAW couplings and does not recompute the cavity at each magnetic field orientation.

The shallow-groove PnC is designed using a 2D plane-strain Bloch FEM of the Ge/SiGe/Si stack.
For each sampled groove geometry, the calculation tracks the Rayleigh-like surface branches near the Brillouin-zone edge and identifies a Rayleigh-like surface stop band centered at $2$~GHz, rather than a complete elastic band gap.
The Bloch attenuation is then transferred to a finite-mirror Bragg coupled-mode model that contains the selected left and right mirror cells, the material loss, and the monitored right output port.
This model returns the loaded quality factor $Q_{\text{L}}$, the right-port output fraction $\beta_{\text{R}}$, and the effective longitudinal mode length $L_{\text{eff}}$ collected in Table~\ref{tab:pnc_design}.

The total loaded linewidth decomposes as:
\begin{equation}
\label{Eq:AppendixKappaDecomposition}
\begin{aligned}
\kappa &= \frac{\omega_c}{Q_{\text{L}}}=\kappa_{\text{R}}+\kappa_{\text{L}}+\kappa_{\text{int}} \,, \\
\beta_{\text{R}} &= \frac{\kappa_{\text{R}}}{\kappa} \,,
\end{aligned}
\end{equation}
where $\kappa_{\text{R}}$ and $\kappa_{\text{L}}$ are the decay rates through the right and left mirrors and $\kappa_{\text{int}}$ is the intrinsic material loss.
Thus $\beta_{\text{R}}\simeq0.95$ means that approximately $95\%$ of the loaded cavity decay leaves through the monitored right port in the model, before detector inefficiency is included.
The transverse confinement remains an explicit engineering input, so this is a device-anchored hybrid design estimate rather than a direct 3D defect-eigenmode zero-point-strain calculation.
\begin{table}[htbp!]
\caption{Selected parameters of the $2$~GHz shallow-groove PnC defect cavity.}
\label{tab:pnc_design}
\begin{ruledtabular}
\begin{tabular}{lcc}
Quantity & Symbol & Value \\
\colrule
PnC period                    & $a_{\text{PnC}}$       & $785.9$~nm \\
Groove width                  & $w_{\text{g}}$           & $235.8$~nm \\
Groove depth                  & $d_{\text{g}}$           & $45$~nm \\
Surface-stop-band width       & $\Delta f_{\text{gap}}$ & $64.3$~MHz \\
Left mirror cells             & $N_{\text{L}}$           & $93$ \\
Right mirror cells            & $N_{\text{R}}$           & $48$ \\
Defect length                 & $L_{\text{d}}$           & $1.983$~$\mu$m \\
Loaded quality factor         & $Q_{\text{L}}$           & $4016$ \\
Effective mode length         & $L_{\text{eff}}$        & $11.07\lambda_{\text{PnC}}$ \\
Right-port output fraction    & $\beta_{\text{R}}$       & $0.950$ \\
\end{tabular}
\end{ruledtabular}
\end{table}
$\lambda_{\text{PnC}}=2\pi/k_c$ is the carrier wavelength associated with the selected defect mode.

The FEM-LKBP projection yields the complex transverse matrix elements:
\begin{equation}
\begin{aligned}
v_{\xi}^{\perp}(\theta_B) &= \mel*{1}{V^{\text{C}}_{\xi,+}}{0} \,, \\
\xi &= \pm1 \,,
\end{aligned}
\end{equation}
introduced in Appendix~\ref{App:RayleighMode}.
These matrix elements are technical intermediates: after energy normalization they define the co-rotating quantum coupling $g^{\text{co}}_{\xi k}$ of Sec.~\ref{Sec_4_spin_phonon}, while their flux-normalized magnitudes define the open-SAW rates $\Gamma_\xi(\theta_B)$ and the total scale $\Gamma_{\text{open}}(\theta_B)=\Gamma_+(\theta_B)+\Gamma_-(\theta_B)$.

The transverse focusing factor $F_{\text{focus}}$, the effective longitudinal mode length $L_{\text{eff}}$, and the velocity used in the conversion are instead engineering and model inputs.
The angle-dependent angular-frequency coupling scale is given by the following expression:
\begin{equation}
g_0^2(\theta_B)=\Gamma_{\text{open}}(\theta_B)F_{\text{focus}}v_{\text{SAW}}/L_{\text{eff}} \,.
\end{equation}
Here $\Gamma_{\text{open}}(\theta_B)$ carries the total microscopic strength of the spin-phonon coupling for the open launcher.
The factor $F_{\text{focus}}$ accounts for the transverse compression of the acoustic field that the 2D model does not resolve.
The factor $v_{\text{SAW}}/L_{\text{eff}}$ converts a continuum emission rate into the free spectral range of a single longitudinal cavity mode. The calculation takes the open-SAW phase velocity $v_{\text{SAW}}$.
We take $F_{\text{focus}}=W_{\text{open}}/W_{\text{eff}}=12/1.6=7.5$, where $W_{\text{open}}=12~\mu\mathrm{m}$ is the open-SAW aperture and $W_{\text{eff}}=1.6~\mu\mathrm{m}$ is the assumed effective transverse cavity width.

The local coupling of qubit $j$ to the selected defect mode is reconstructed as:
\begin{equation}
g_j(\theta_B,x_j)=g_0(\theta_B)\mathcal{C}_j(\theta_B,x_j) \,,
\end{equation}
with the dimensionless local interference factor given by:
\begin{equation}
\mathcal{C}_j(\theta_B,x_j)=\frac{v_{+}^{\perp}(\theta_B)\mathrm{e}^{\mathrm{i}kx_j}+\mathrm{e}^{\mathrm{i}\phi_c}v_{-}^{\perp}(\theta_B)\mathrm{e}^{-\mathrm{i}kx_j}}{\sqrt{2\left(|v_{+}^{\perp}(\theta_B)|^2+|v_{-}^{\perp}(\theta_B)|^2\right)}} \,.
\end{equation}
The normalized factor $\mathcal{C}_j$ retains the direction-resolved relative amplitudes, phases, and position dependence, while $g_0(\theta_B)$ carries the total microscopic coupling strength through $\Gamma_{\text{open}}(\theta_B)$.
Thus the herald-dark interval originates from the suppression of the total coupling scale near the transverse-coupling node, while chirality controls the standing-wave contrast.

A global phase is removed so that $g_1$ is real and positive, while $g_2/g_1$ retains the physical relative phase.
The cavity phase is calibrated to give constructive interference at qubit 1 for $\theta_B^{\text{ref}}=36^\circ$ and is then held fixed throughout the angular scan.
Qubit 1 is fixed at $x_1=0$, while the placement coordinate $s=2(x_2-x_1)/\lambda_{\text{PnC}}\bmod 1$ sets $x_2$.
Because the transverse confinement is imposed rather than solved for, the resulting $g_j$ constitute a device-anchored hybrid estimate rather than a self-consistent 3D prediction.

The two qubits and the defect mode are evolved explicitly from $\ket{11,0_c}$.
Cavity leakage through the right mirror is the monitored jump channel, while leakage through the remaining ports, thermal absorption, qubit relaxation, and pure dephasing are unmonitored.
The first registered phonon defines the heralding event; its timestamp is used for a local phase correction, and a second registered phonon during the feedback latency rejects the trajectory.

The cavity mode is retained explicitly, and no adiabatic elimination or phenomenological Purcell rate is used.
In a frame rotating at the cavity frequency, the coherent Hamiltonian is:
\begin{equation}\label{Eq:AppendixBellHamiltonian}
\frac{H}{\hbar} = \sum_{j=1}^2\Delta_j\sigma_j^+\sigma_j^-+\sum_{j=1}^2(g_j\sigma_j^+\hat{a}+g_j^*\hat{a}^\dagger\sigma_j^-) \,.
\end{equation}
Here $\hat{a}$ annihilates a cavity phonon, while $\sigma_j^-=\ketbra{0}{1}_j$ and $\sigma_j^+=\ketbra{1}{0}_j$ act on the Zeeman doublet of qubit $j$.
The detunings $\Delta_j$ and couplings $g_j$ in Eq.~\eqref{Eq:AppendixBellHamiltonian}, as well as all rates below, are angular frequencies.
The final Fig.~\ref{Fig_6} calculation uses $\Delta_1=\Delta_2=0$ and a cavity Hilbert-space cutoff $n_c=3$, which retains the states with zero, one, and two phonons.

The total cavity linewidth $\kappa=\omega_c/Q_{\text{L}}$ decomposes into the port rates of Eq.~\eqref{Eq:AppendixKappaDecomposition}, with $\kappa_{\text{R}}=\beta_{\text{R}}\kappa$, and the thermal occupation is $n_{\text{th}}=[\mathrm{e}^{hf_c/(k_{\text{B}}T_{\text{bath}})}-1]^{-1}$.
The monitored right-port jump and the two unmonitored cavity jumps are:
\[
\begin{gathered}
L_h=\sqrt{\eta_{\text{det}}\kappa_{\text{R}}(n_{\text{th}}+1)}\hat{a}, \\
L_u=\sqrt{(\kappa-\eta_{\text{det}}\kappa_{\text{R}})(n_{\text{th}}+1)}\hat{a}, \\
L_T=\sqrt{\kappa n_{\text{th}}}\hat{a}^\dagger.
\end{gathered}
\]
Here $L_u$ represents unregistered cavity emission, while $L_T$ represents thermal absorption.
Qubit relaxation and pure dephasing are represented by the following unmonitored jumps:
\[
\begin{gathered}
L_{1,j}=\sqrt{\gamma_1}\sigma_j^-,\quad L_{\phi,j}=\sqrt{\gamma_\phi/2}\sigma_j^z.
\end{gathered}
\]
The rates entering these qubit jump operators are $\gamma_1=1/T_1$ and $\gamma_\phi=1/T_2-1/(2T_1)$. $T_2$ denotes the effective exponential dephasing time used in the Markovian Lindblad model.

At $T_{\text{bath}}=4$~mK and $f_c=2$~GHz, $n_{\text{th}}$ is negligible but is retained in the solver.
The open-SAW continuum is used only to set $g_0$, and its residual branching factor $r_{\text{open}}$ is fixed to zero to avoid counting the same reservoir twice.
The final calculation further assumes ideal detection, $\eta_{\text{det}}=1$, $r_d=0$, and $t_{\text{reset}}=0$, so the reported fidelity and rate are upper bounds with respect to the detection chain.
Accordingly, the receiver in Fig.~\ref{Fig_1}(b) is represented only by an ideal time-resolved detection channel; no microscopic transduction or detector-response model is included.

For any unmonitored jump $L$, the dissipator is $D[L]\rho=L\rho L^\dagger-\{L^\dagger L,\rho\}/2$.
The unnormalized no-click state obeys:
\begin{equation}\label{Eq:AppendixBellNoClick}
\begin{aligned}
\dv{\rho_0}{t} =& -\frac{\mathrm{i}}{\hbar}[H,\rho_0]+D[L_u]\rho_0-\frac{1}{2}\{L_h^\dagger L_h,\rho_0\}\\
&+\sum_{j=1}^2(D[L_{1,j}]\rho_0+D[L_{\phi,j}]\rho_0) +D[L_T]\rho_0\,.
\end{aligned}
\end{equation}
Let $K_0$ denote the trace-decreasing generator on the right-hand side of Eq.~\eqref{Eq:AppendixBellNoClick}, and let $J_h[\rho]=L_h\rho L_h^\dagger$ denote the registered-phonon jump.
The nonzero counted blocks propagated for Fig.~\ref{Fig_6} obey:
\begin{align}
\dv{\rho_0}{t} &= K_0\rho_0 \,, \\
\dv{\rho_h}{t} &= J_h[\rho_0]\label{Eq:AppendixBellRhoH} \,, \\
\dv{\rho_1}{t} &= K_0\rho_1+J_h[\rho_0] \,.
\end{align}
The initial conditions are $\rho_0(0)=\ketbra{11,0_c}{11,0_c}$ and $\rho_h(0)=\rho_1(0)=0$.
The absorbing block $\rho_h$ stores the state at the first registered phonon, whereas $\rho_1$ retains trajectories with exactly one registered click while continuing the no-click evolution.

For timestamp bin $b$ with boundaries $t_b$ and $t_{b+1}$, the first-click contribution stored in the absorbing block of Eq.~\eqref{Eq:AppendixBellRhoH} is $\Delta\rho_b=\rho_h(t_{b+1})-\rho_h(t_b)$.
Each bin is propagated for the feedback latency $\tau_{\text{fb}}$ with $K_0$, so any second registered phonon during that latency removes trace and rejects the attempt.
An ideal timestamp-dependent local $Z$ operation $U_b$ removes the dynamical phase accumulated up to the click time of bin $b$.
Every accepted bin is then aligned with the same microscopic target phase of the Bell state:
\[
\begin{gathered}
\ket{\Psi_B(\phi_B)}=[\ket{01}+\mathrm{e}^{\mathrm{i}\phi_B}\ket{10}]/\sqrt{2}.
\end{gathered}
\]
The phase $\phi_B=-\arg(g_2/g_1)$ is fixed by the microscopic coupling ratio rather than optimized from the accepted state.

The unnormalized accepted state is $\rho_{\text{acc}}=\sum_bU_b\mathrm{e}^{K_0\tau_{\text{fb}}}\Delta\rho_bU_b^\dagger$.
This procedure distinguishes the click-time resolution $\delta t_{\text{click}}$ from the physical feedback latency $\tau_{\text{fb}}$.
The former sets the timestamp binning, while a second click during the latter rejects the attempt.
The procedure applies no separate preparation, storage, feedback-success, or readout map.
The accepted state is evaluated immediately after the timestamp correction and feedback latency.
The subsequent detuning from the cavity described in Sec.~\ref{Sec_5_2_bell} is not implemented as a separate dynamical map.

The PnC geometry, the cavity phase, $Q_{\text{L}}$, $\beta_{\text{R}}$, $L_{\text{eff}}$, the detector assumptions, and the feedback parameters remain fixed throughout the scan of Fig.~\ref{Fig_6}.
At each $\theta_B$, only the precomputed complex matrix elements $v_{\pm}^{\perp}(\theta_B)$ and their flux-normalized rate scale are updated.
No cavity parameter is reoptimized with $\theta_B$, and the cavity phase remains fixed at the value calibrated at $\theta_B^{\text{ref}}=36^\circ$.
The placement coordinate $s$ changes only the position of qubit 2.

The accepted probability is given by the trace of the unnormalized accepted state, $P_{\text{acc}}=\operatorname{Tr}[\rho_{\text{acc}}]$.
The normalized two-qubit state is obtained by tracing out the cavity, $\rho_q=\operatorname{Tr}_c[\rho_{\text{acc}}]/P_{\text{acc}}$.
The conditional Bell fidelity is evaluated on this state as $F_{\text{Bell}}^{\text{cond}}=\mel*{\Psi_B(\phi_B)}{\rho_q}{\Psi_B(\phi_B)}$.
The mean attempt cycle time is then given by the expression:
\begin{equation}
\bar{t}_{\text{cyc}} = \int_0^{T_h}\operatorname{Tr}[\rho_0(t)] \dd{t}+t_{\text{reset}}+P_{\text{first}}\tau_{\text{fb}} \,.
\end{equation}
Here $P_{\text{first}}=\operatorname{Tr}[\rho_h(T_h)]$ is the probability of a first registered phonon within the heralding window.
The accepted heralding rate follows from the accepted probability and cycle time as $R_{\text{acc}}=P_{\text{acc}}/\bar{t}_{\text{cyc}}$.
For each scan point, the target phase is $\phi_B(\theta_B,s)=-\arg[g_2(\theta_B,s)/g_1(\theta_B,s)]$; it is fixed by the microscopic couplings at that point and is not optimized from $\rho_q$.
The fixed cavity, open-system, feedback, and numerical inputs are summarized in Table~\ref{tab:app_lindblad_parameters}.

The grey band of Fig.~\ref{Fig_6} marks the contiguous near-dark angular interval around $90^\circ$ for which $\max_s P_{\text{acc}}(\theta_B,s)<10^{-4}$.
At $(\theta_B,s)=(36^\circ,0)$, the calculation gives $F_{\text{Bell}}^{\text{cond}}\simeq0.88$ and $R_{\text{acc}}\simeq82~\mathrm{s}^{-1}$.
The inverse rate, $R_{\text{acc}}^{-1}\simeq12~\mathrm{ms}$ at the reference point, is the mean wall-clock waiting time per accepted state under the assumed zero reset time; it is not the duration of the successful click-conditioned evolution.
\begin{table}[htbp!]
\caption{Fixed parameters used for the timestamp-corrected Bell-state calculation in Fig.~\ref{Fig_6}.}
\label{tab:app_lindblad_parameters}
\begin{ruledtabular}
\begin{tabular}{cc}
Parameter & Value \\\hline
$f_c,Q_{\text{L}},\beta_{\text{R}}$ & $2~\mathrm{GHz},4016.11,0.950079$ \\
$L_{\text{eff}}/\lambda_{\text{PnC}},F_{\text{focus}}$ & $11.0659,7.5$ \\
$\kappa/(2\pi)$ & $0.497994~\mathrm{MHz}$ \\
$T_{\text{bath}},T_1,T_2$ & $4~\mathrm{mK},10~\mathrm{ms},10~\mu\mathrm{s}$ \\
$\eta_{\text{det}},r_d,r_{\text{open}}$ & $1,0,0$ \\
$T_h,\delta t_{\text{click}},\tau_{\text{fb}}$ & $100~\mu\mathrm{s},1~\mu\mathrm{s},1~\mu\mathrm{s}$ \\
$t_{\text{reset}},n_c$ & $0,3$ \\
$\Delta_1/(2\pi),\Delta_2/(2\pi)$ & $0,0~\mathrm{MHz}$ \\
$\theta_B^{\text{ref}},N_\theta,N_s$ & $36^\circ,181,120$ \\
$F_{\text{Bell}}^{\text{cond}},R_{\text{acc}}$ at $(36^\circ,0)$ & $0.8802,81.7497~\mathrm{s}^{-1}$ \\
\end{tabular}
\end{ruledtabular}
\end{table}

\section{High-coupling cascaded-entanglement benchmark}
\label{App:CascadeBenchmark}
\begin{figure}[htbp!]
\centering
\includegraphics[width=\linewidth]{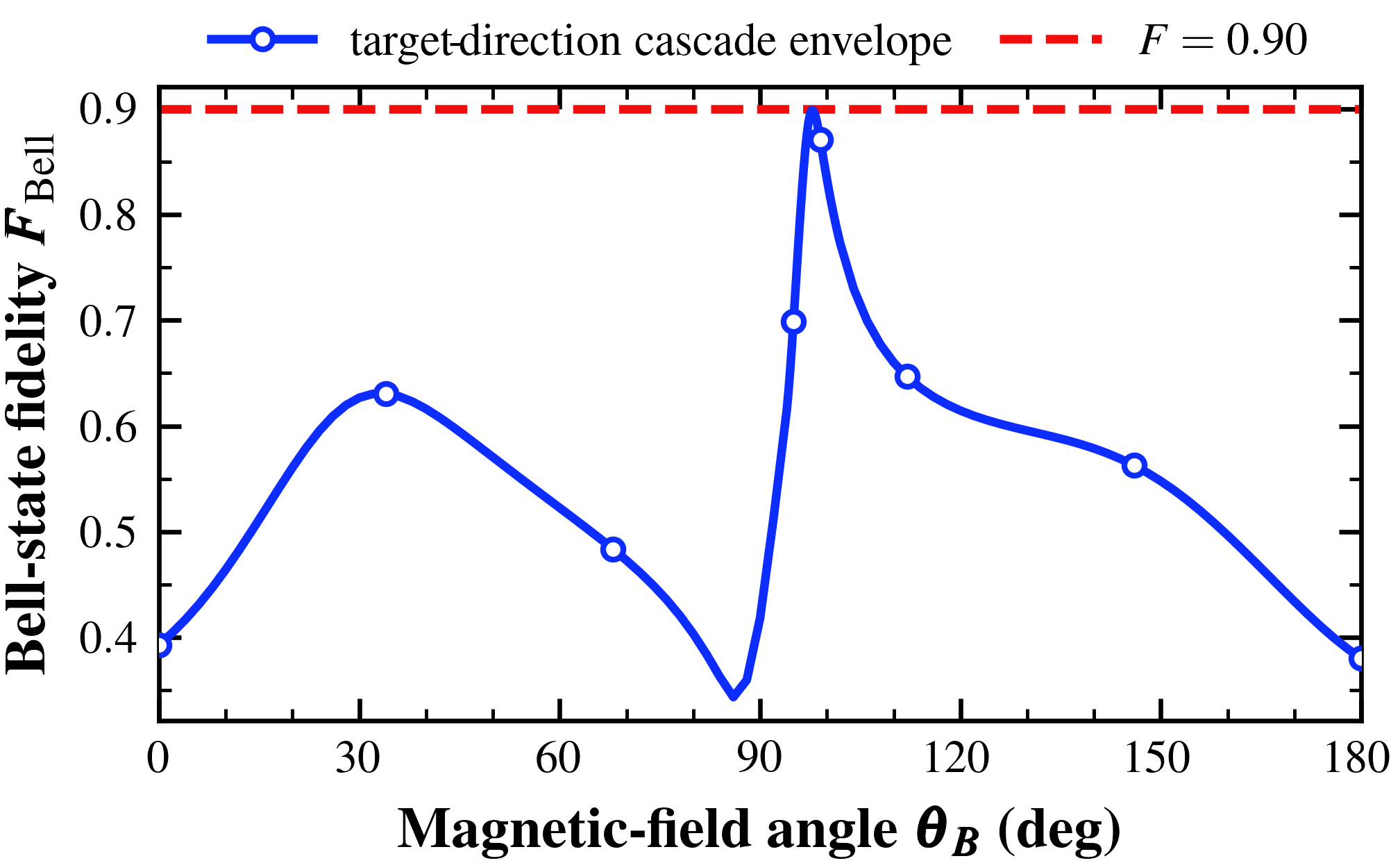}
\caption{Target-direction cascaded-entanglement benchmark.
Optimized steady-state Bell fidelity $F_{\mathrm{Bell}}$ as a function of magnetic-field angle $\theta_B$.
At each angle, the two qubits are locally trimmed to their respective $2$~GHz resonances, while the common Rabi amplitude and opposite detuning are reoptimized.
The useful right-going channel is treated as the collective cascade, whereas reverse-direction emission is retained as independent loss.
The blue curve reaches $F_{\mathrm{Bell}}=0.8988$ at $\theta_B=97.75^\circ$; the red dashed line marks the reference level $F_{\mathrm{Bell}}=0.90$.}
\label{Fig:AppendixCascade}
\end{figure}
This appendix considers a prospective cascaded-entanglement benchmark for a future transversely confined, quasi-1D phononic waveguide operated in a slow-wave regime.
The FEM-LKBP calculation supplies the magnetic field dependence, relative phases, and directional branching ratios of the co-rotating spin-phonon couplings.
The absolute waveguide coupling scale and group velocity are imposed as benchmark parameters (not derived from the present open-SAW or PnC structures).

The two qubits are driven with equal Rabi amplitudes and opposite detunings, while their emission into the selected propagation direction forms a collective jump channel.
We write the local direction-resolved emission operators as $c_j=\sqrt{\Gamma_{j,+}}\mathrm{e}^{\mathrm{i}\phi_j}\sigma_-^{(j)}$ for each qubit $j$.
Here $\sigma_-^{(j)}=\ketbra{0}{1}_j$ is the lowering operator in the Zeeman basis of Appendix~\ref{App:QubitProjection}.
The rate $\Gamma_{j,+}$ describes emission into the useful right-going channel, and $\phi_j$ collects the propagation and coupling phases.
We describe the target-direction channel by the collective jump operator $L_+=c_1+c_2$ and the cascaded interaction:
\begin{equation}
H_{\text{casc}}=\frac{\hbar}{2\mathrm{i}} \left(c_2^\dagger c_1-c_1^\dagger c_2\right) \,.
\end{equation}

The driven bright component radiates into this common channel, whereas coherent downstream reabsorption stabilizes an odd-parity dark steady state.
Equal Rabi amplitudes and optimized opposite detunings then maximize its Bell-state component.
The selected right-going channel is treated as the useful cascade.
The intended role of chirality is to reduce coupling to the reverse direction, but the treatment of the reverse channel and the degree of its suppression must be specified by the complete master equation below.

At each $\theta_B$, the angular dependence of the direction-resolved co-rotating coupling rates and their directional branching ratios is retained.
These quantities come from the same FEM and six-band qubit projection used throughout this work.
The two qubits are locally trimmed to their respective $2$~GHz resonances.
The common Rabi amplitude and opposite detuning are then optimized at each angle.

The two-qubit basis is ordered as $\{\ket{00},\ket{01},\ket{10},\ket{11}\}$, with qubit 1 upstream of qubit 2.
In the frame rotating at the common drive frequency, the local Hamiltonian is:
\begin{equation}
\frac{H_{\text{loc}}}{\hbar} = \frac{\Omega}{2}\sum_{j=1}^{2}\left(\sigma_j^+ + \sigma_j^-\right) + \frac{\delta}{2}\left(\sigma_1^z - \sigma_2^z\right) \,.
\end{equation}
With $\sigma_j^z=\ketbra{1}{1}_j-\ketbra{0}{0}_j$, the two rotating-frame detunings are $\Delta_1=+\delta$ and $\Delta_2=-\delta$.

The reverse-direction emission operators are $r_1=\sqrt{\Gamma_{1,-}}\sigma_1^-$ and $r_2=\sqrt{\Gamma_{2,-}}\sigma_2^-$, which enter as independent loss channels.
The master equation used for the reported target-direction calculation is:
\begin{equation}\label{Eq:AppendixCascadeMasterEquation}
\begin{aligned}
\dv{\rho}{t} =& -\frac{\mathrm{i}}{\hbar}[H_{\text{loc}}+H_{\text{casc}},\rho]+(n_{\text{th}}+1)D[L_+]\rho \\
&+ n_{\text{th}}D[L_+^\dagger]\rho+\sum_{j=1}^{2}[(n_{\text{th}}+1)D[r_j]\rho \\
& +n_{\text{th}}D[r_j^\dagger]\rho+D[L_{1,j}]\rho+D[L_{\phi,j}]\rho] \,,
\end{aligned}
\end{equation}
where $D[L]\rho=L\rho L^\dagger-\{L^\dagger L,\rho\}/2$, $L_{1,j}=\sqrt{\gamma_1}\sigma_j^-$, $L_{\phi,j}=\sqrt{\gamma_\phi/2}\sigma_j^z$, $\gamma_1=1/T_1$, and $\gamma_\phi=\max\{0,1/T_2-1/(2T_1)\}$.
The FEM separation is $x_2-x_1=4\lambda_{\text{wg}}=6.681~\mu\mathrm{m}$, where $\lambda_{\text{wg}}$ is the wavelength used in the waveguide benchmark.
This does not conflict with the smaller benchmark group velocity used below, because the latter represents a strongly dispersive slow-wave operating point and determines the one-dimensional phonon density of states.
The calculation employs the zero-delay Markov approximation, so this separation does not appear as an explicit propagation-delay parameter in Eq.~\eqref{Eq:AppendixCascadeMasterEquation}.

For a periodic waveguide-normalization length $L_{\text{q}}$, the density of states of each propagation branch per unit angular frequency is $\rho_\xi(\omega)=L_{\text{q}}/(2\pi v_{g,\xi})$, and we have:
\begin{equation}
\Gamma_{j,\xi} = 2\pi\abs{g_{j,\xi}^{\text{wg}}}^{2}\rho_\xi(\omega_q) = \frac{L_{\text{q}}}{v_{g,\xi}}\abs{g_{j,\xi}^{\text{wg}}}^{2} \,,
\end{equation}
where $b_{j,\pm}=[1\pm\chi_j(\theta_B)]/2$, $\abs{g_{j,\pm}^{\text{wg}}}^{2}=b_{j,\pm}\abs{g_j^{\text{wg}}}^{2}$, and $\abs{g_j^{\text{wg}}}^{2}=\abs{g_{j,+}^{\text{wg}}}^{2}+\abs{g_{j,-}^{\text{wg}}}^{2}$.
The box-normalized amplitudes scale as $L_{\text{q}}^{-1/2}$, whereas the physical emission rates $\Gamma_{j,\xi}$ are independent of normalization length.
For the prospective benchmark, we choose $L_{\text{q}}=1~\mu\mathrm{m}$.
The externally imposed benchmark values $g_1^{\text{wg}}/(2\pi)=g_2^{\text{wg}}/(2\pi)=2.11$~MHz, and equal group velocity $352$~m/s, correspond to $\Gamma_{1,\text{1D}}=\Gamma_{2,\text{1D}}=0.5$~MHz, respectively~\cite{Modica2020,Yu2020}.

These externally imposed waveguide benchmark parameters are distinct from the PnC cavity couplings $g_j$ of Appendix~\ref{App:BellStateDynamics}.
They define a prospective slow-wave regime in which the directional emission rate is sufficiently large to compete with the intrinsic qubit decoherence.
With the coherence and thermal parameters specified above, the optimized steady-state fidelity reaches $F_{Bell} = 0.8988$ near $\theta_{B} = 97.75^\circ$.
The narrow angular window reflects the simultaneous requirements of strong directional selectivity, matched coupling to the useful cascaded channel, and optimized coherent driving.

Cascaded stabilization of this kind is the acoustic analogue of the driven-dissipative schemes developed for chiral photonic networks~\cite{Pichler2015,Zou2022}.
The benchmark demonstrates the performance that could become accessible after substantial waveguide confinement and directional-isolation engineering.
It should not be compared quantitatively with the PnC heralding rate or conditional fidelity, because the two calculations use different architectures, coupling scales, and figures of merit.

\clearpage

\end{document}